\documentclass[letterpaper,10pt]{article}
\usepackage{graphicx}
\usepackage{opex3}
\usepackage{amsmath,amssymb}

\newcommand{\myfig}[2]{
    \begin{figure}[!h]
    \begin{centering}
    \includegraphics[width=0.8\textwidth]{#1}
    \caption{#2}\label{#1}
    \end{centering}
    \end{figure}
}

\newcommand{\Fig}[1]{Fig.~\ref{#1}}
\newcommand{\eq}[1]{eq.~\eqref{#1}}
\newcommand{\fig}[1]{fig.~\ref{#1}}
\newcommand{\BI}{\begin{itemize}\item}
\newcommand{\I}{\item}
\newcommand{\EI}{\end{itemize}}
\newcommand{\BEAS}{\begin{eqnarray}}
\newcommand{\EEAS}{\end{eqnarray}}
\newcommand{\BSA}{\begin{subequations}\begin{align}}
\newcommand{\ESA}{\end{align}\end{subequations}}
\newcommand{\BE}{\begin{equation*}}
\newcommand{\EE}{\end{equation*}}
\newcommand{\BIT}{\begin{itemize}}
\newcommand{\EIT}{\end{itemize}}

\DeclareMathOperator*{\minimize}{\text{minimize}\quad}
\newcommand{\subto}{\text{subject to}\quad}
\newcommand{\T}{^\dagger}

\begin{document}

\title{Nanophotonic Computational Design}
\author{Jesse Lu$^\ast$ and Jelena Vu\v{c}kovi\'{c}}
\address{Stanford University, Stanford, California, USA.}
\email{jesselu@stanford.edu}

\begin{abstract}
In contrast to designing nanophotonic devices 
    by tuning a handful of device parameters, 
    we have developed a computational method 
    which utilizes the full parameter space to design linear nanophotonic devices.
We show that our method may indeed be capable of designing 
    any linear nanophotonic device by demonstrating designed structures which
    are fully three-dimensional and multi-modal,
    exhibit novel functionality,
    have very compact footprints,
    exhibit high efficiency, and
    are manufacturable.
In addition, we also demonstrate the ability to produce structures
    which are strongly robust to wavelength and temperature shift,
    as well as fabrication error.
Critically, we show that our method 
    does not require the user to be a nanophotonic expert or 
    to perform any manual tuning. 
Instead, we are able to design devices 
    solely based on the user's desired performance specification for the device.
\end{abstract}
\ocis{230.75370, 130.3990.}

\section{Introduction}
Currently, almost all nanophotonic components are designed 
    by hand-tuning a small number of parameters 
    (e.g. waveguide widths and gaps, hole and ring sizes).
However, the realization of 
    increasingly complex, dense, and robust on-chip optical networks
    will require utilizing increasing numbers of parameters
    when designing nanophotonic components.

Opening the design space to include many more parameters
    allows for smaller footprint, higher performance devices by definition,
    since original designs are still included in this parameter space.
Unfortunately, the lack of intuition for what such designs might look like and
    the inability to manually search such a large parameter space
    have greatly hindered the ability to achieve this.

For this reason, we have developed and implemented a computational method
    which is able to use the full parameter space 
    to design linear nanophotonic components in three dimensions.
Critically, our method requires no user intervention or manual tuning.
Instead, a \emph{design-by-specification} scheme is used 
    to produce designs based solely on a user's performance specification.

We show that our method can indeed produce designs 
    which are extremely compact, and, at the same time, highly efficient.
Furthermore, we demonstrate that devices with novel functionality
    are easily designed.
We also show that our method can be used to produce designs
    with extreme robustness to wavelength and temperature shift,
    as well as fabrication error.

Lastly, all our results are produced by simply specifying
    the functionality and performance of the desired device,
    which suggests that our method may indeed by able
    to design \emph{all} linear nanophotonic devices.

\section{Problem formulation}
In order to produce designs which utilize the full parameter space,
    and are based solely on the user's performance specification,
    we formulate the design problem in the following way:
\begin{subequations}\begin{align}
    \minimize & \sum_i^M \|A_i(z)x_i - b_i\|^2 \\
    \subto & \alpha_{ij} \le |c_{ij}\T x_i| \le \beta_{ij}, \quad
        \text{for $i = 1, \ldots, M$ and $j = 1, \ldots, N_i$} \\
        &   z_\text{min} \le z \le z_\text{max}
\end{align} \label{problem}\end{subequations}

The explanation for the various terms in \eq{problem} follows:
\begin{enumerate}
\item 
    $A_i(z)x_i - b_i$ is the \emph{physics residual} for the $i$th mode.
    That is to say, $A_i(z)x_i - b_i$ represents the underlying physics
        of the problem; namely, the electromagnetic wave equation
        \mbox{$(\nabla\times\mu_0^{-1}\nabla\times - \omega_i^2 \epsilon) E_i 
            +i \omega_i J_i $}.

    The specific substitutions used in order to transform
        \BE (\nabla\times\mu_0^{-1}\nabla\times - \omega_i^2 \epsilon) E_i 
            +i \omega_i J_i  \quad\longrightarrow\quad A_i(z)x_i - b_i \EE
        are
    \BI $E_i \to x_i$,
    \I  $\epsilon \to z$,
    \I  $\nabla\times\mu_0^{-1}\nabla\times - \omega_i^2 \epsilon \to A_i(z)$, and
    \I  $ -i\omega_i J_i \to b_i$.  \EI

    In contrast to typical schemes for optimizing physical structures,
        our formulation actually allows for non-zero physics residuals;
        which can be deduced since $A_i(z)x_i-b_i=0$ is not a hard constraint.
    Instead, this formulation is what we call an \emph{objective-first} \cite{ob1_wg}
        formulation in that the \emph{design objective} (explained below)
        is prioritized above satisfying physics.

\item
    The (field) design objective consist of 
        the constraint $\alpha_{ij} \le |c_{ij}\T x_i| \le \beta_{ij}$.
    Physically, this constraint describes 
        the performance specification of the device 
        via a series of field overlap integrals 
        at various output ports of the device.
    Specifically, the $c_{ij}\T x_i$ terms represents an overlap integral between
        the E-field of the $i$th mode ($x_i$)
        with an E-field of the user's choice ($c_{ij}$),
        where the additional subscript $j$ allows the user
        to include multiple such fields.
    The amplitude of the overlap integral is then forced to reside between
        $\alpha_{ij}$ and $\beta_{ij}$.

    This mechanism allows the user to express 
        the desired performance of the device
        as a combination of field amplitudes in various output field patterns.
    These outputs would be in response to a predefined input excitation,
        which is determined by the current excitation $b_i$ ($-i\omega_i J_i$)
        in the physics residual of each mode.

    As an example of a design objective for some mode 1 
        a user might choose to have the majority of the output power
        reside in some output pattern 1,
        while ensuring that only a small amount of power 
        be transferred to some output pattern 2.
    In this case the user would use 
        $0.9 \le |c_{11}\T x_1| \le 1.0$ for the former.
        and then $0.0 \le |c_{12}\T x_1| \le 0.01$ for the latter;
        where $c_{11}$ and $c_{12}$ are representative of 
        output patterns 1 and 2 respectively.
        
    Finally, we note again that the design objective in our formulation
        is actually a hard constraint.
    This means that it is \emph{always satisfied}, 
        even to the extent of allowing for an unphysical field 
        (since the physics residual will not be exactly 0).
    It is for this reason that we call such a formulation ``objective-first''.

\item 
    The final term in \eq{problem}, $z_\text{min} \le z \le z_\text{max}$,
        is the structure design objective.
    It is used as a relaxation of the binary constraint,
        $z \in \{z_\text{min}, z_\text{max}\}$,
        which would ensure that the final design be composed 
        of two discrete materials.
\end{enumerate}

\section{Method of solution}
We employed the alternating directions method of multipliers (ADMM) algorithm 
    \cite{admm}
    in order to solve \eq{problem}.
The ADMM algorithm solves \eq{problem} by iteratively solving for 
    $x_i$, $z$, and a dual variable $u_i$.

Since we are working in three dimensions, solving \eq{problem} for $x_i$ 
    is non-trivial in that it involves millions of variables and
    requires solving for the ill-conditioned $A_i(z)$ matrix.
For this reason, we use a home-built
    finite-difference frequency-domain (FDFD) solver which 
    implements a hardware-accelerated iterative solver\cite{wonseok}
    on Amazon's Elastic Compute Cloud.
Critically, our cloud-based solver allows us to scale to solve problems
    with arbitrarily-large number of modes,
    with no significant penalty in runtime.

In contrast to solving for $x_i$, solving for $z$ is much simpler since we
    only consider planar structures;
    thereby limiting $z$ to have only thousands of variables.

Lastly, in order to arrive at fully discrete, manufacturable structures,
    we convert $z$ to a boundary parameterization\cite{levelset}
    and tune our structure using a steepest-descent method.

\section{Results}
We demonstrate the effectiveness of our design method 
    by producing designs for a variety of nanophotonic devices.

All of our results are in three dimensions
    and are planar structures, consisting of a 250 nm etched silicon slab
    completely surrounded by silica.
The permittivity values of silicon and silica used
    were $\epsilon_\text{Si} = 12.25$ and $\epsilon_\text{SiO$_2$} = 2.25$
    respectively.

Many, if not all, of the produced designs exhibit 
    novel functionality, high efficiency, and 
    very compact footprints of only a few square vacuum wavelengths,
    while still remaining manufacturable.
We also show that many devices can be designed
    to exhibit different functionality for different input excitations.
Additionally, we show that devices can be designed with large tolerances for
    errors in wavelength, temperature and fabrication.

\subsection{Mode converters}

Our first devices consisted of waveguide mode converters.
Such devices are simple in that they are single-input, single-output devices.
At the same time, such a device is significant because 
    it demonstrates the feasibility of multi-mode on-chip optical networks
    by showing that high-efficiency mode conversion 
    can readily be achieved in planar on-chip nanophotonic structures.

We show, through the design of mode converters for both the TE- and TM-polarized
    waveguide modes, 
    that our method is indeed fully three-dimensional.
Additionally, the devices also demonstrate very small device footprints 
    ($1.6 \times 2.4$ microns for this device in particular
    operating at 1550 nm wavelength).

\subsubsection{TE mode converter}
Our first result is a mode conversion device operating in the TE polarization,
    where the primary E-field component
    of the waveguide mode is polarized in the plane of the structure.

Our performance specification (\fig{3Db_wgc_te2}) 
    for the device was for $\ge90\%$ of the 
    input power to be transferred from the fundamental waveguide mode,
    to the second-order waveguide mode.
At the same time, we specified that no more than 1\% of the input power
    was to remain in the transmitted fundamental mode.

\myfig{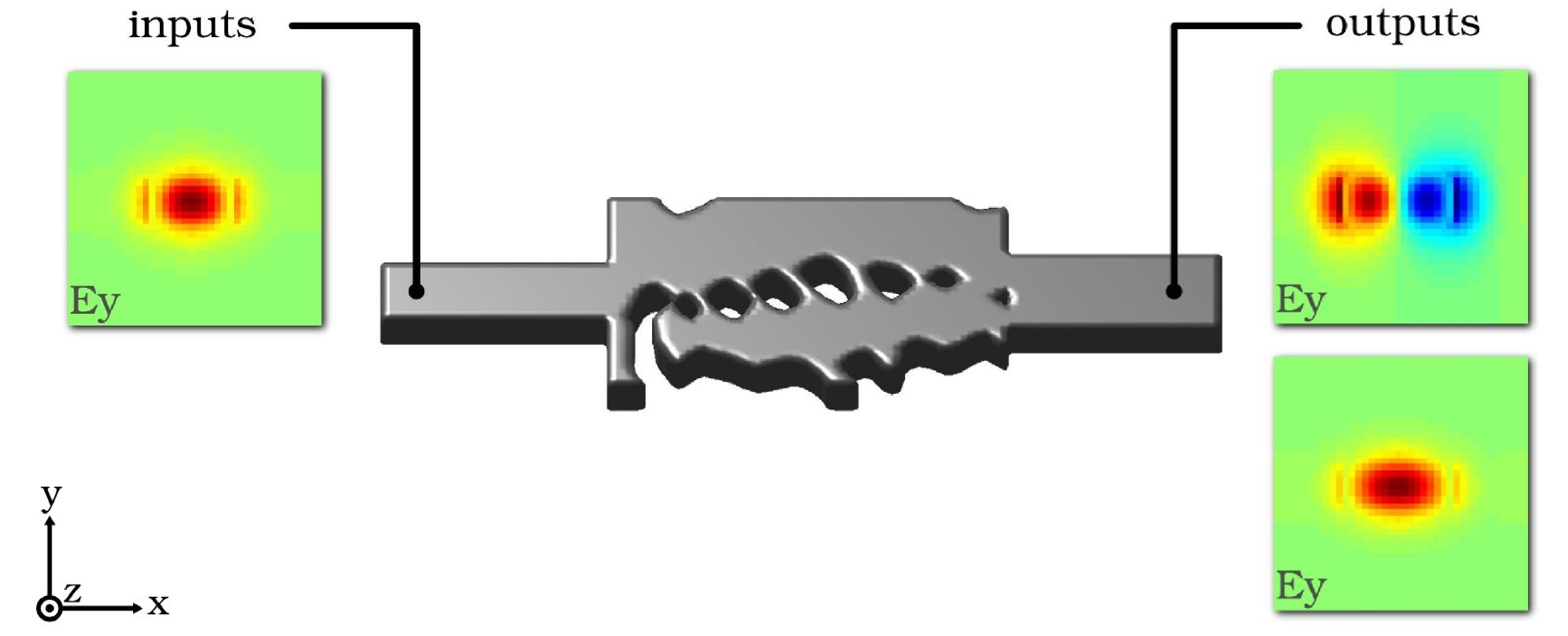}
    {Perfomance specification of the TE mode converter.
    Input mode is the fundamental TE-polarized mode on the left.
    Primary output mode is the second-order mode on the right.
    Output power in the transmitted fundamental mode 
    should be no more than 1\%.
    The structure shown is the final three-dimensional design
    (the same holds for all the following figures in the article).
    }

The performance of the device is shown in \fig{takeaway}.
The conversion efficiency into the second-order mode is lower
    than desired (86.4\%). 
Imperfect conversion may be due to evanescent modes ``interfering'' 
    with the output field overlap calculation.

\myfig{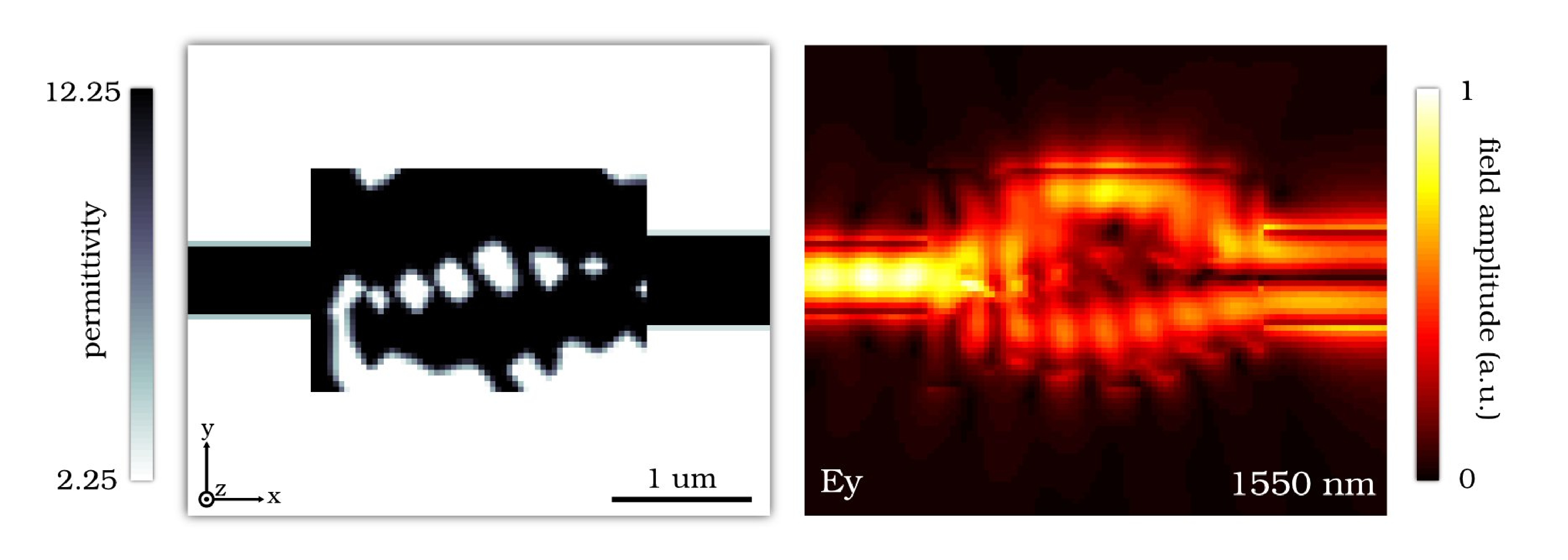}
    {Structure and E-field at the central plane of the TE mode converter.
    The conversion efficiency into the second-order mode is 86.4\%,
    while the power into the rejection mode (fundamental) is 0.7\%.
    Device footprint is $1.6\times2.4$ microns.
    Operating wavelength is 1550 nm.}

\subsubsection{TM mode converter}

In addition to mode conversion in the TE polarization (E-field in-plane),
    we show that TM polarization (E-field out-of-plane) mode converters
    can be designed as well.
This example shows that full three-dimensional structures 
    truly are possible,
    and that no approximations are needed for our method.

Since our method is design-by-specification, 
    the design of a TM mode converter requires only
    a small modification to the performance specification of the device;
    namely the polarization of the input and output modes (\fig{3Db_wgc_tm}).
Specifically, we still design for 
    $\ge 90\%$ conversion into the second-order mode and 
    a $\le 1\%$ allowance for the fundamental mode to be transmitted.

\myfig{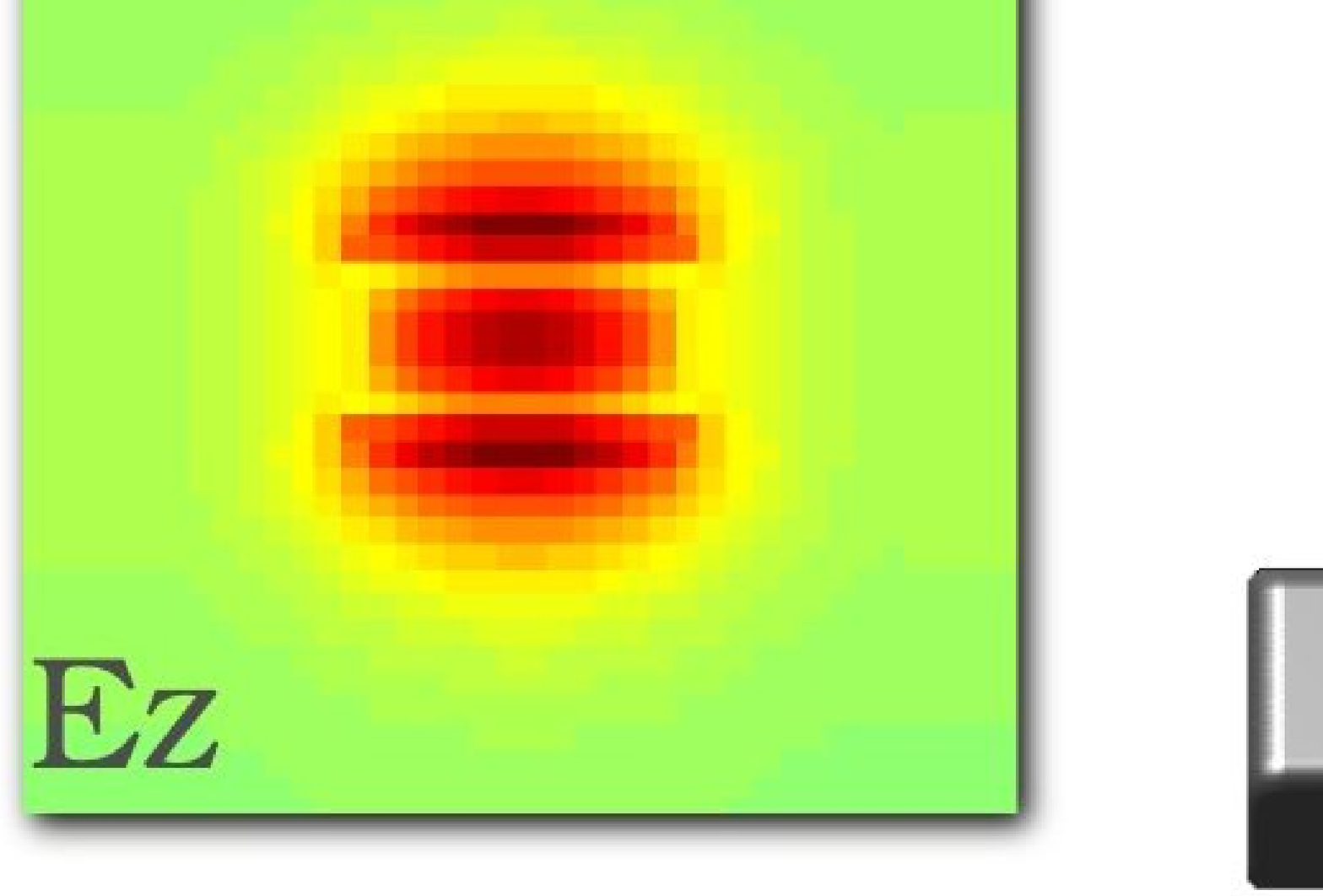}
    {Perfomance specification of the TM mode converter.
    Input mode is the fundamental TM-polarized mode on the left.
    Primary output mode is the second-order mode on the right.
    Output power in the transmitted fundamental mode on the right 
    above 1\% is to be rejected.
    The structure shown is the final three-dimensional design.}
    
The performance of the device is shown in \fig{wgc_tm}.
The lower conversion efficiency of 76.9\% in contrast to the TE mode converter
    may be attributed to the lower confinement of the TM waveguide modes 
    in such thin slabs.
However, good rejection of only 1\% is still achieved.

\myfig{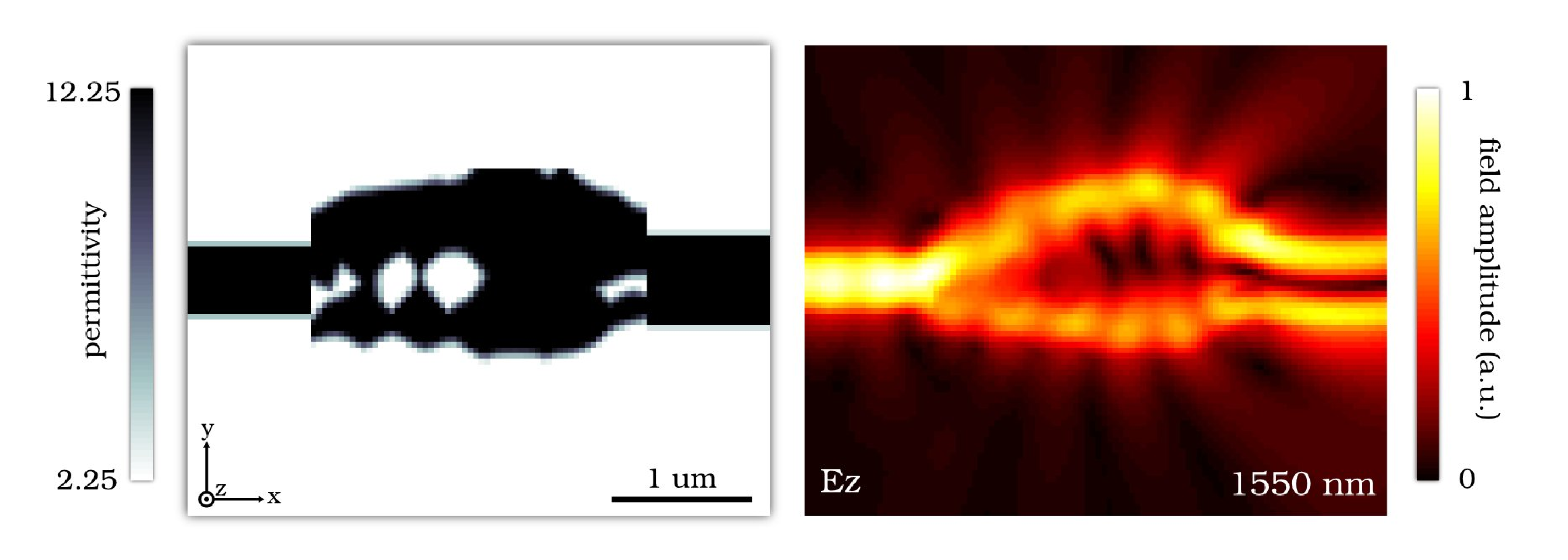}
    {Structure and E-field at the central plane of the TM mode converter.
    The conversion efficiency into the second-order mode is 76.9\%,
    while the power into the rejection mode (fundamental) is 1.0\%.
    Device footprint is $1.6\times2.4$ microns.}

\subsection{Mode splitters}
Next, we demonstrate the design of nanophotonic waveguide mode splitters.
Such devices can be used as multiplexers or demultiplexers and
    are the key component in utilizing a single waveguide to transmit 
    multiple optical signals.

As a demonstration of the versatility of our method, 
    we show that it is capable of designing mode splitting devices 
    based on either the spatial profile, the polarization, or the wavelength
    of the input modes.

The performance specification for each device is simply to convert
    more than 90\% of the input power in a particular input mode into
    either one of the output modes.
At the same time, we specify that the transmission into the other output mode 
    be kept below 1\% of input power.

\subsubsection{Spatial mode splitter}
We demonstrate what is, to our knowledge, 
    the first design for a three-dimensional
    nanophotonic spatial mode splitter 
    (previous designs were restricted to two dimensions \cite{2dsplitter}).
Such a device is the key enabler for multi-mode on-chip optical circuits,
    and we show here that they can be designed to be highly efficient
    while utilizing a very small device footprint ($2.8\times2.8$ microns). 
The performance specification is shown in \fig{3Db_spl_modal},
    and the final results is shown in \fig{spl_modal}.

\myfig{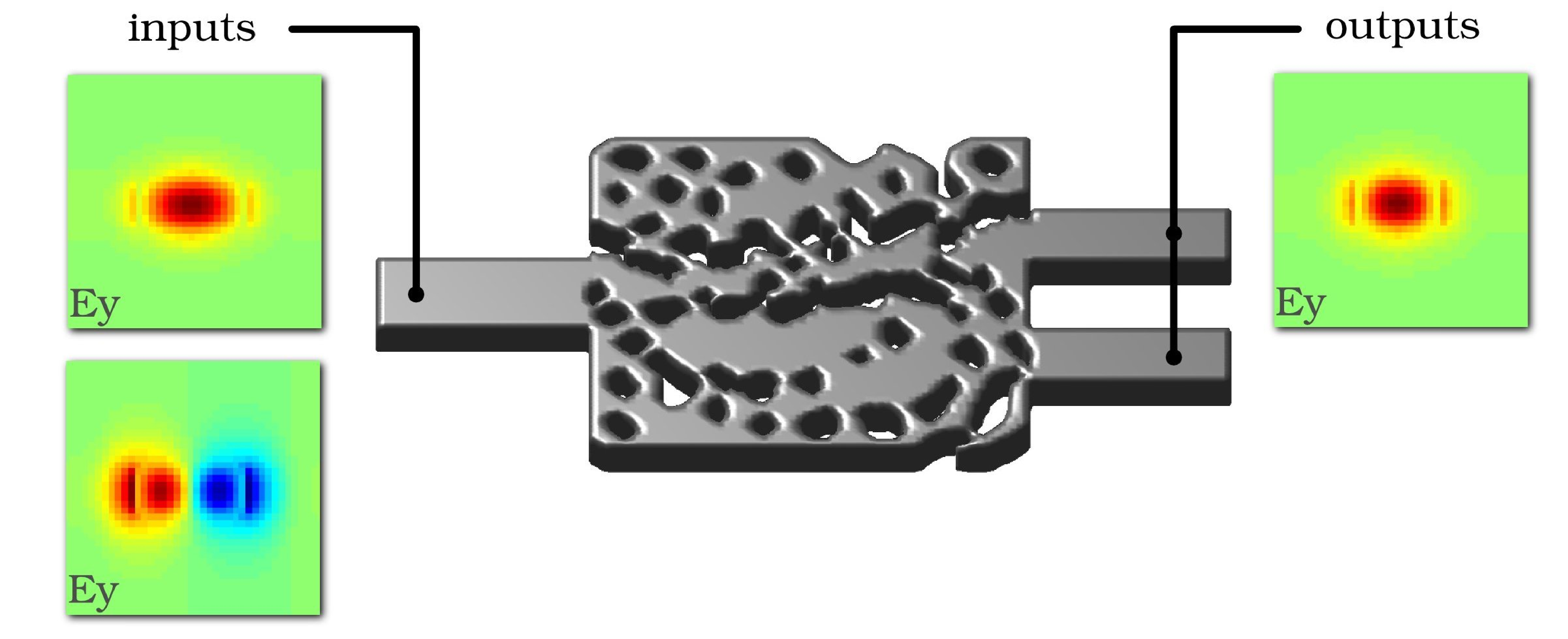}
    {Spatial mode splitter performance specification.
    Input mode is either the fundamental or second-order
        TE-polarized mode on the left.
    Output modes are the fundamental waveguide modes of either output 
        waveguide on the right.
    Output power into the desired output arm is specified to be greater than 90\%,
        while power into the opposing arm is set to below 1\%.}
\myfig{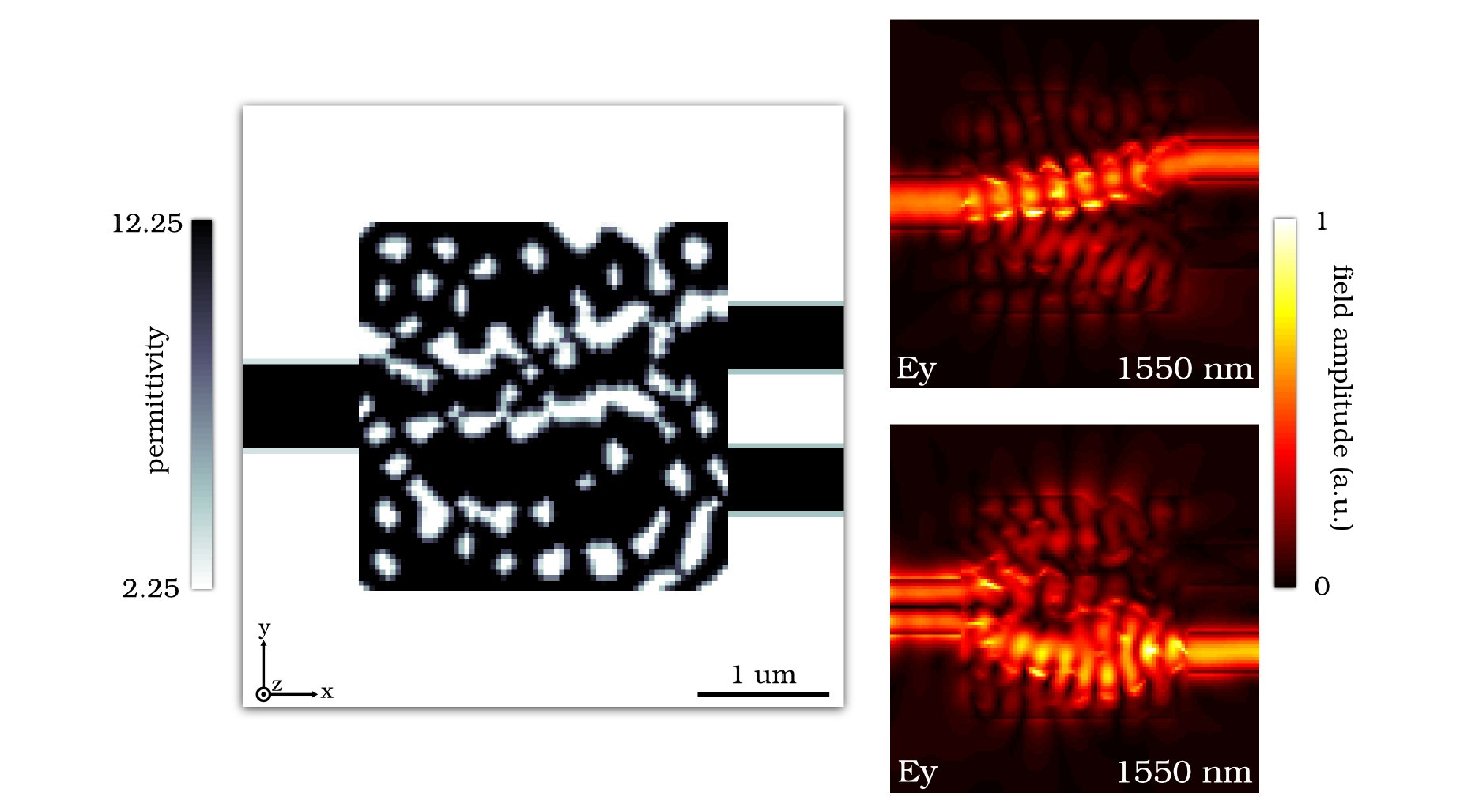}
    {Spatial mode splitter final result.
    The conversion efficiencies into the upper and lower output arms
        are 88.7\% and 77.4\% respectively, 
        while the rejection powers for the same modes are 0.27\% and 0.20\%.
    Device footprint is $2.8\times2.8$ microns.}

\subsubsection{TE/TM splitter}
In addition to splitting different spatial modes, 
    we show that different polarizations can also be split.
\Fig{3Db_spl_tetm} shows the performance specification of a 
    device which is able to separate fundamental TE-polarized ($E_y$ dominant)
    and TM-polarized ($E_z$ dominant)
    waveguide modes into separate arms.
The final, verified result is shown in \fig{spl_tetm}.

Not only is this result the first of its kind,
    it is the first in the device category where a single device
    is able to control both polarizations within the same device footprint.
This shows the versatility and broad applicability of our method.

\myfig{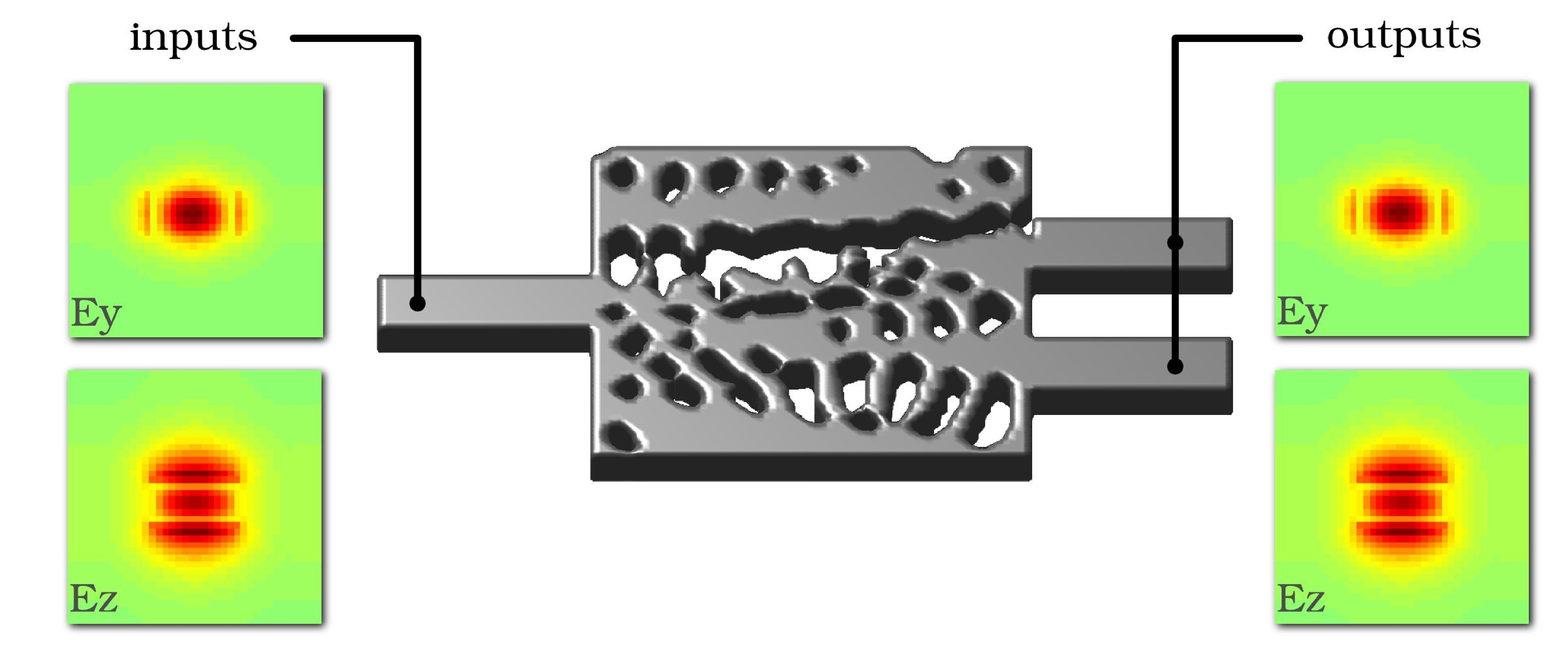}
    {TE/TM splitter performance specification.
    Input mode is either the fundamental TE-polarized ($E_y$ dominant, top left)
    and TM-polarized ($E_z$ dominant, top right).
    Output power into the desired output arm is specified to be greater than 90\%,
        while power into the opposing arm is set to below 1\%.}
\myfig{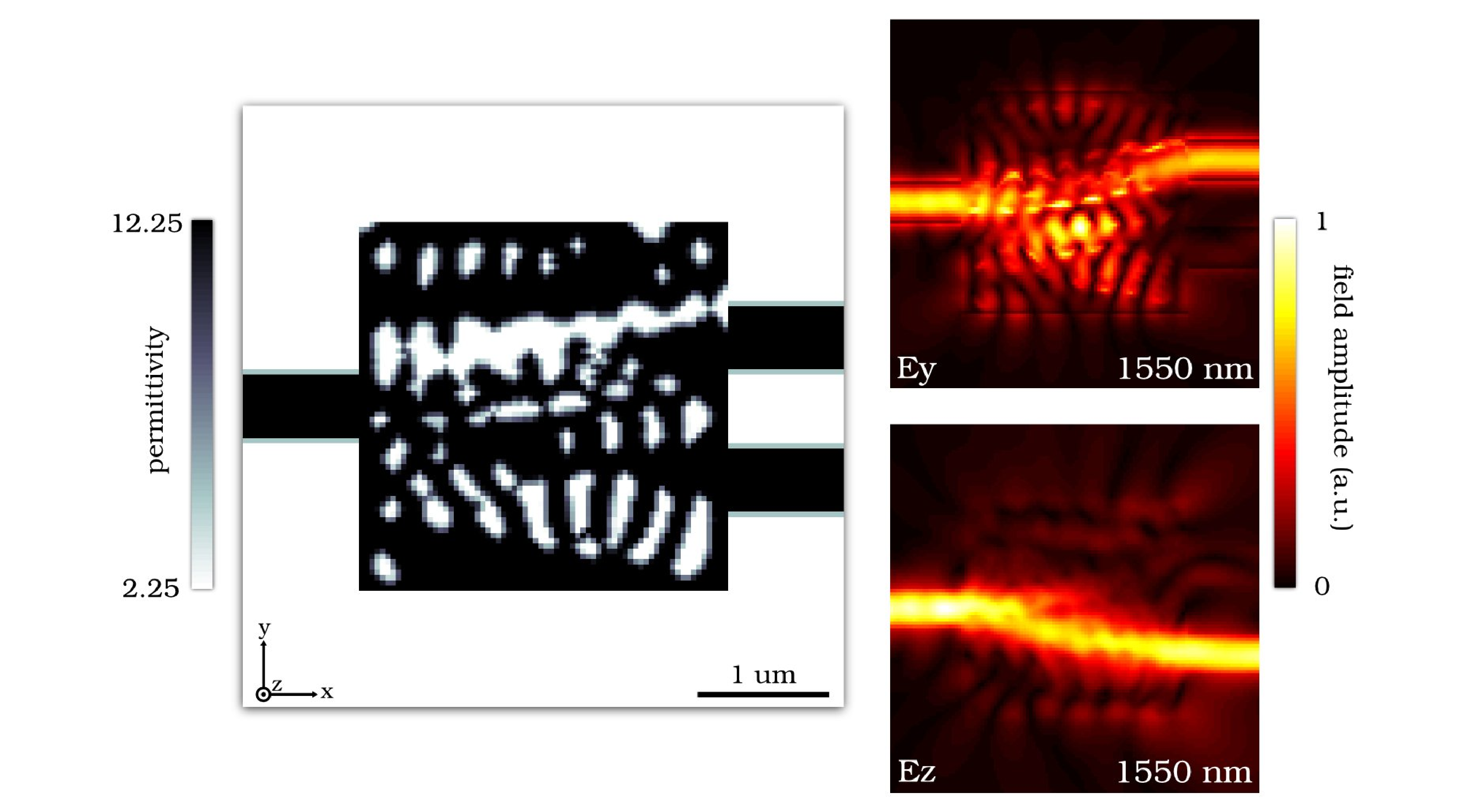}
    {TE/TM splitter final result.
    The conversion efficiencies into the upper and lower output arms
        are 87.6\% and 88.8\% respectively, 
        while the rejection powers for the same modes are 1.06\% and 0.58\%.
    Device footprint is $2.8\times2.8$ microns.}

\subsubsection{Wavelength splitter}
Traditional wavelength splitting devices can also be designed using our method.
Here, we show that the 1550 nm and 1310 nm wavelengths can be split
    in a very small device footprint ($2.8\times2.8$ microns).
The performance specification is shown in \fig{3Db_spl_wdm}, 
    and the final result is shown in \fig{spl_wdm}.
\myfig{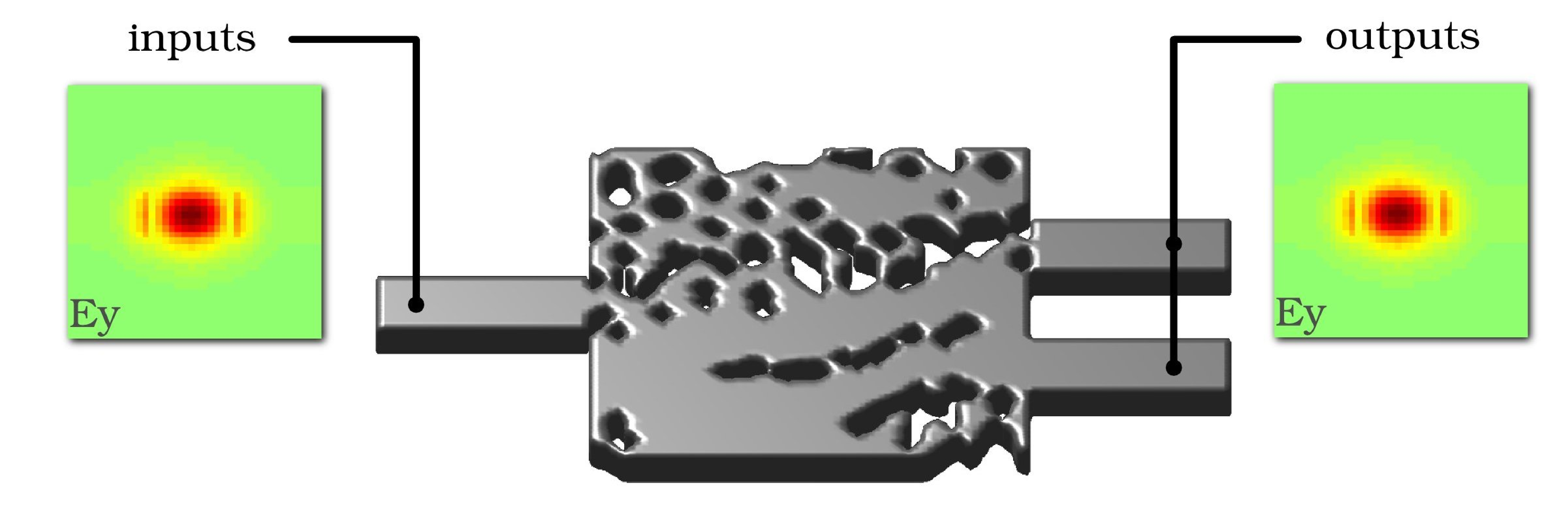}
    {Wavelength splitter performance specification.
    Input mode is the fundamental TE-polarized mode on the left at 
        a wavelength of either 1550 nm or 1330 nm.
    Output modes are the fundamental waveguide modes of either output 
        waveguide on the right;
        however, the 1550 nm wavelength is directed into the top output, 
        while the 1310 nm wavelength is directed into the bottom output.
    Output power into the desired output arm is specified to be greater than 90\%,
        while power into the opposing arm is set to below 1\%.}
\myfig{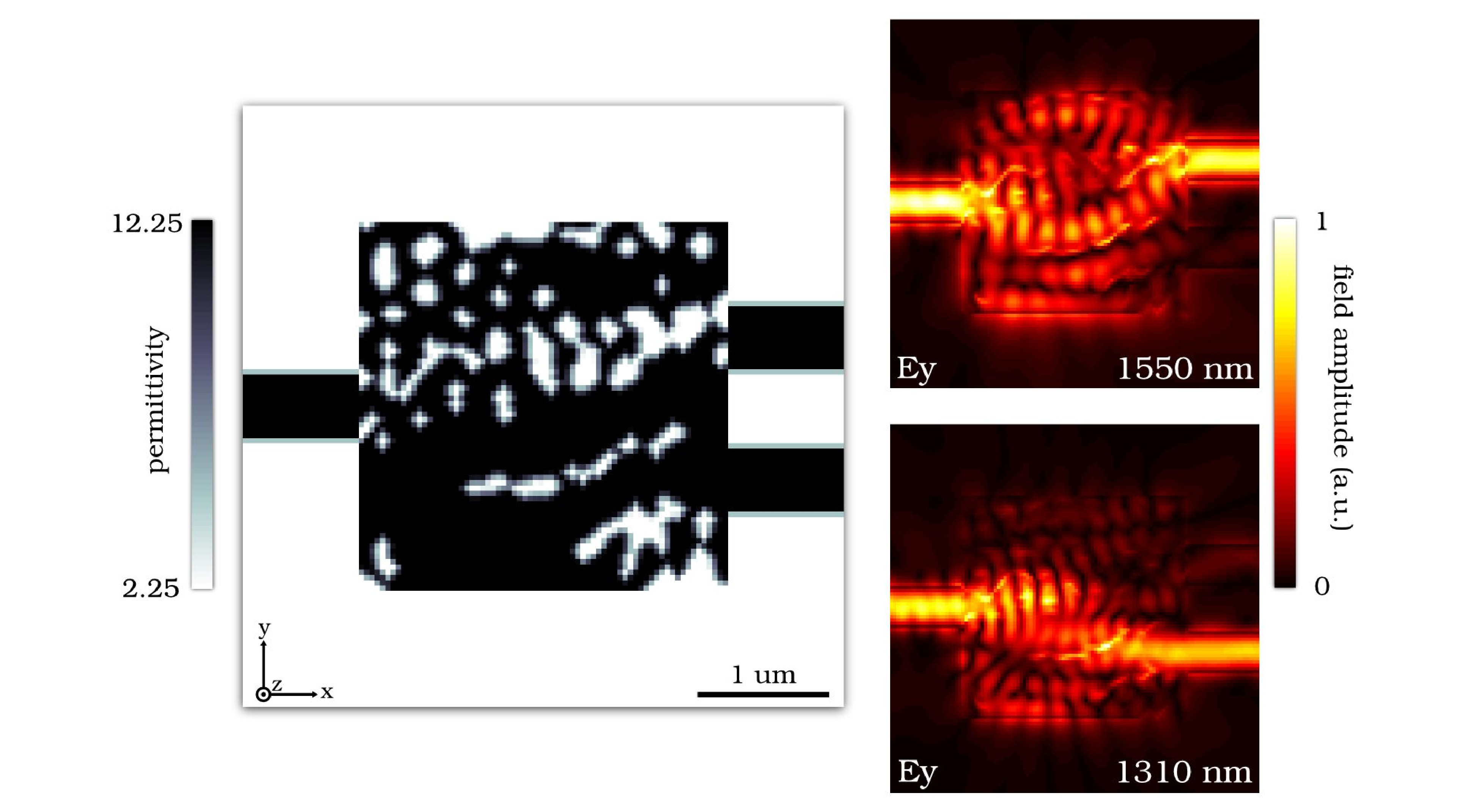}
    {Wavelength splitter final result.
    The conversion efficiencies into the upper and lower output arms
        are 83.2\% and 78.7\% respectively, 
        while the rejection powers for the same modes are 0.49\% and 1.66\%.
    Device footprint is $2.8\times2.8$ microns.}

\subsection{Hubs}
We continue to demonstrate the capabilities of our method
    by designing multi-input, multi-output devices which we call \emph{hubs}.
Such devices essentially re-arrange modes in the waveguides,
    and may be thought of as general cross-connect structures.
Critically, the successful design of such structures 
    shows that efficiently routing overlapping signals can be 
    accomplished in a single layer for nanophotonic circuits.

\subsubsection{3$\times$3 hub}
We first design a hub with three inputs and outputs.
The performance specification is shown in \fig{3Db_hub_3x3},
    and the final result is shown in \fig{hub_3x3}.

\myfig{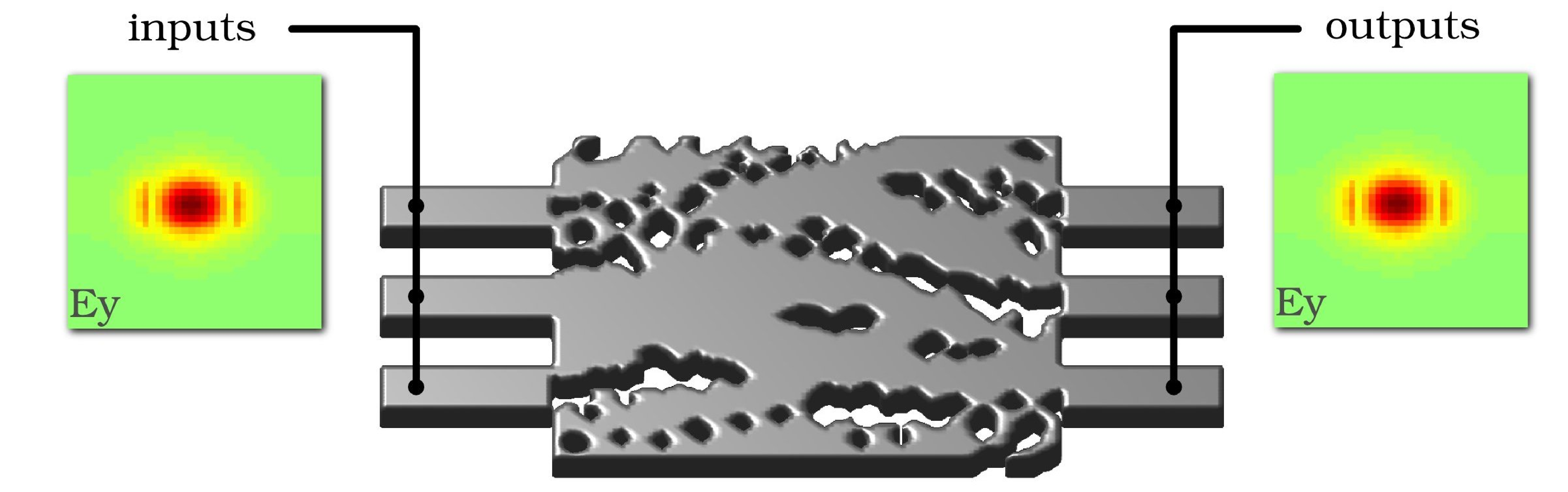}
    {3$\times$3 hub performance specification.
    Input and output modes all consist of the fundamental TE-polarized mode.
    Output power into the desired output arm is specified to be greater than 90\%,
        no rejection modes are used for computational efficiency.
    This hub directs input power from input ports 1, 2, and 3 (from top to bottom)
    into output ports 2, 3, and 1 respectively.}
\myfig{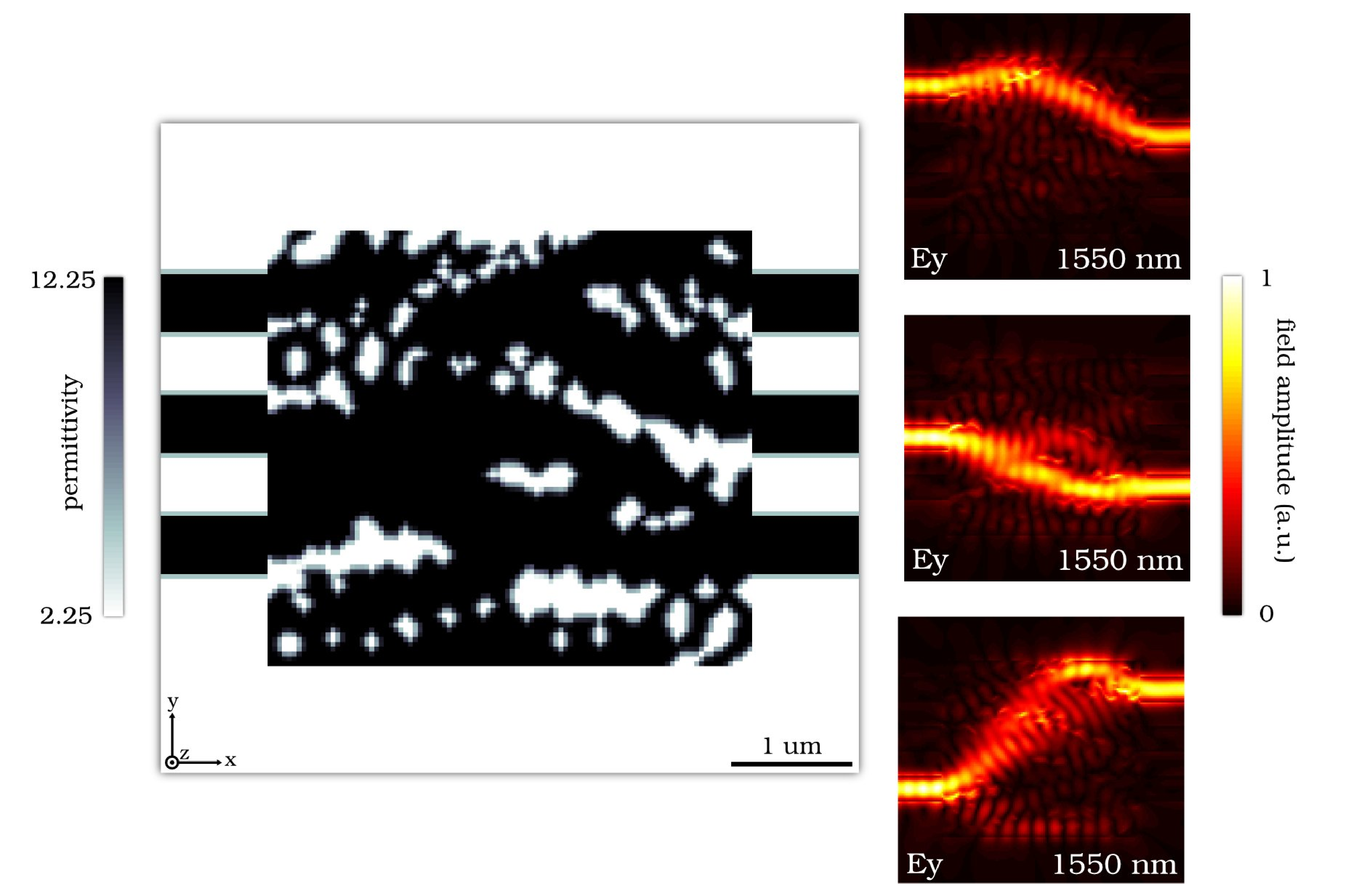}
    {3$\times$3 hub final result.
    The conversion efficiencies into the selected output arms
        are 88.6\%, 90.6\%, and 87.3\% for 
        input arms 1, 2, and 3 respectively (top to bottom).}

\subsubsection{4$\times$4 hub}
We extend our previous result to design a hub with four inputs and outputs.
The performance specification is shown in \fig{3Db_hub_4x4},
    and the final result is shown in \fig{hub_4x4_comb}.

\myfig{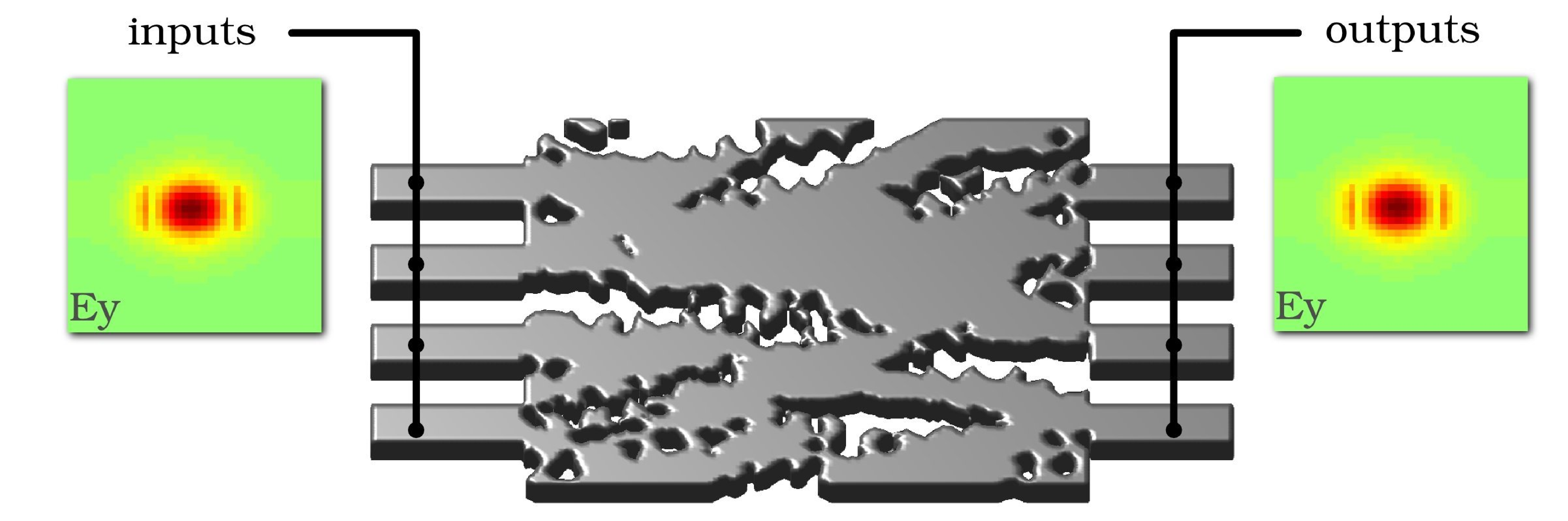}
    {4$\times$4 hub performance specification.
    Input and output modes all consist of the fundamental TE-polarized mode.
    Output power into the desired output arm is specified to be greater than 90\%,
        no rejection modes are used for computational efficiency.
    This hub directs input power from input ports 1, 2, 3, and 4
    into output ports 3, 2, 4, and 1 respectively.}
\myfig{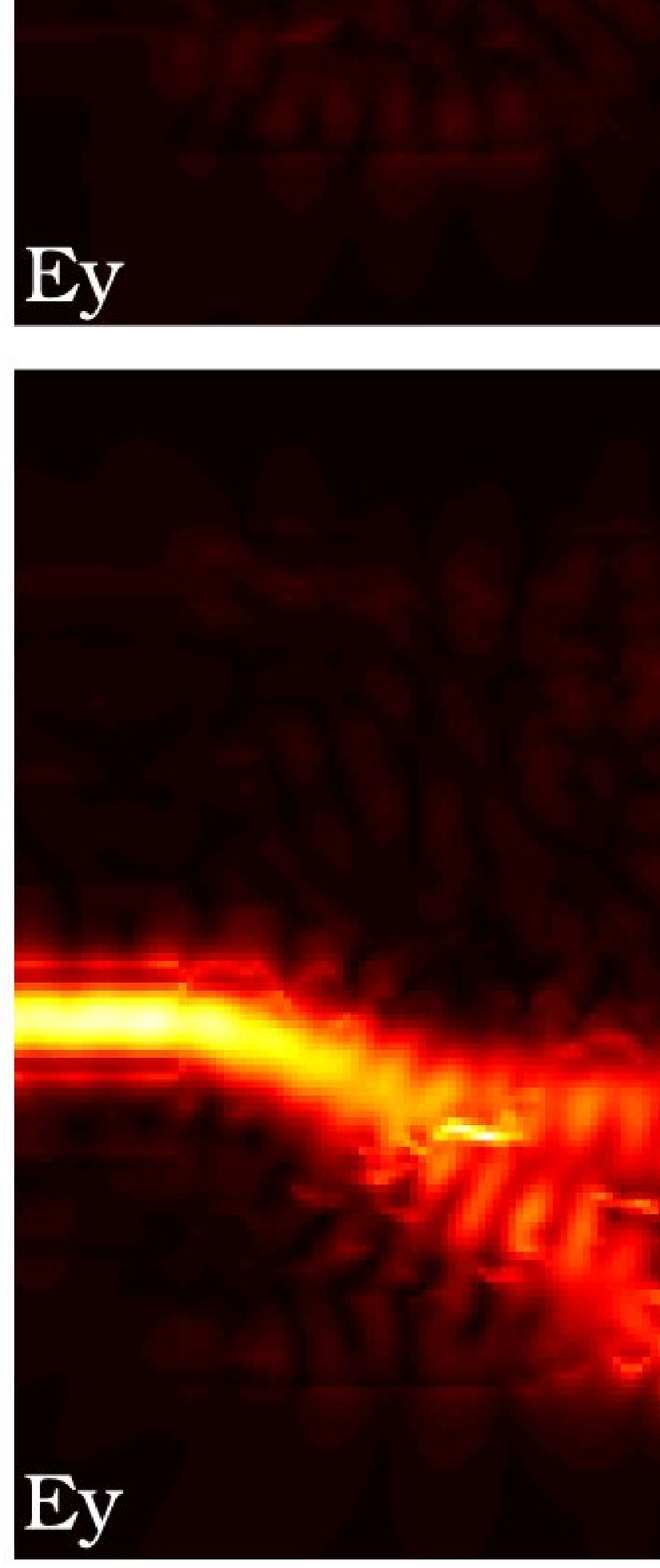}
    {4$\times$4 hub final result.
    The conversion efficiencies into the selected output arms
        are 85.9\%, 88.1\%, 85.4\%, and 84.3\% for
        input arms 1, 2, 3, and 4 respectively (top to bottom).}

\subsubsection{2$\times$2$\times$2 hub}
We can now design a hub that performs 
    different switching functions for different wavelengths.
Specifically, we use two input waveguides, two output waveguides,
    and two wavelengths (hence the name 2$\times$2$\times$2).

Our performance specification (\fig{3Db_hub_2x2x2}) is
    to cross-couple the waveguides at the 1310 nm wavelength,
    but to uncouple the waveguides at the 1550 nm wavelength.
The final result is shown in \fig{hub_2x2x2}.
\myfig{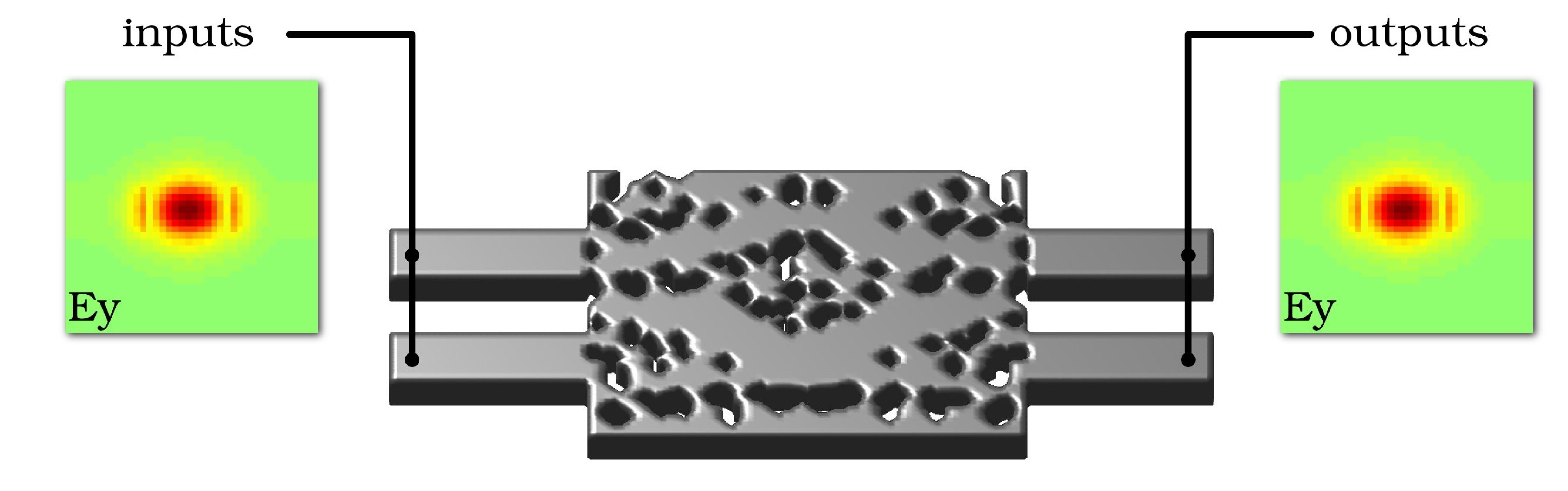}
    {2$\times$2$\times$2 hub performance specification.
    Input and output modes all consist of the fundamental TE-polarized mode
        at either 1550 nm or 1310 nm wavelengths.
    Output power into the desired output arm is specified to be greater than 80\%,
        no rejection modes are used for computational efficiency.
    This hub directs input arms 1 and 2 into output arms 1 and 2 at 1550 nm,
        but swaps them at 1310 nm.}
\myfig{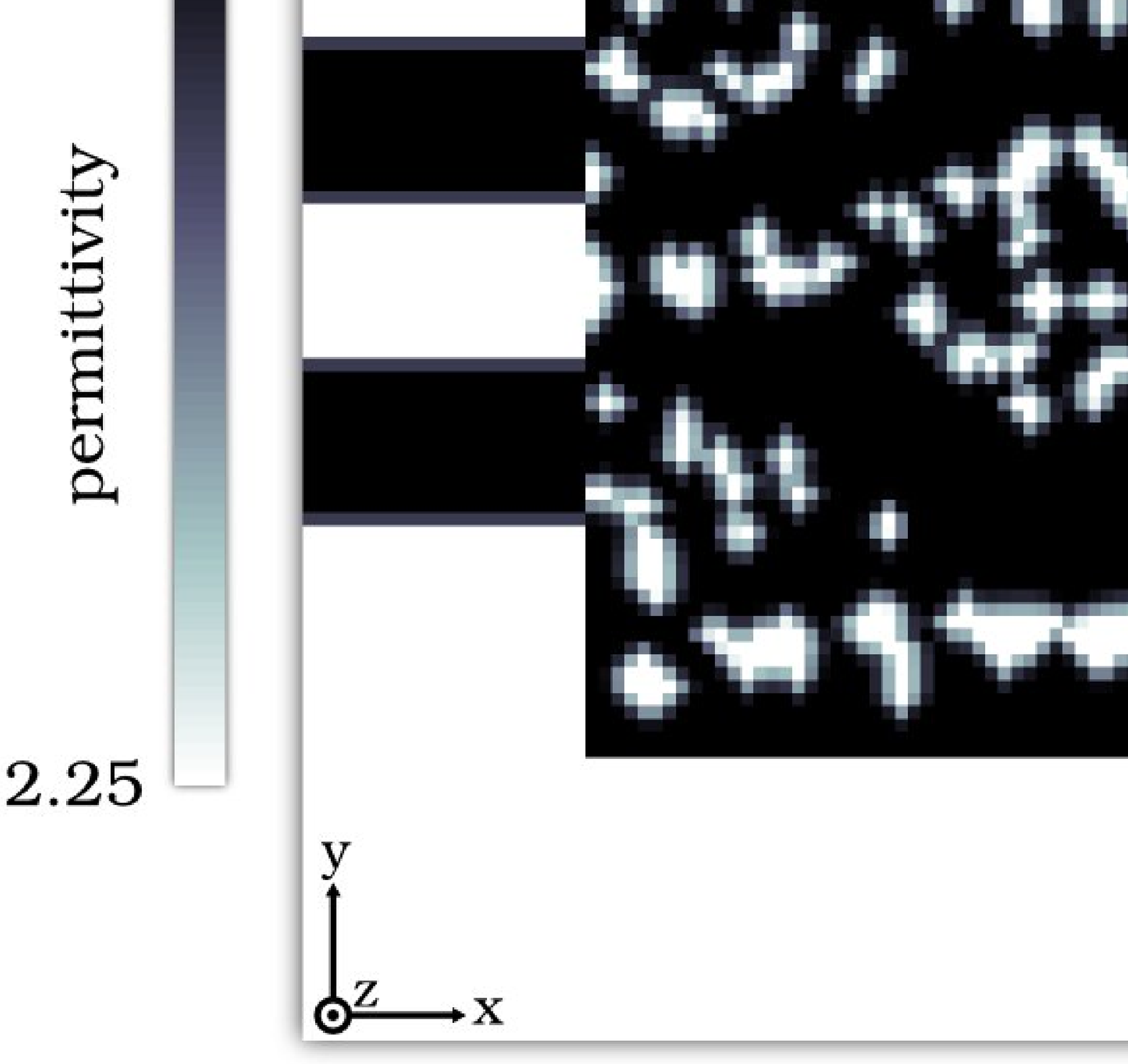}
    {2$\times$2$\times$2 hub final result.
    The conversion efficiencies at the 1550 nm wavelength 
        are 77.6\% and 73.7\% respectively for the top and bottom inputs.
    At 1310 nm, the respective efficiencies are 75.7\% and 75.2\%.}

\subsection{Fiber couplers}
The capabilities of our method are further demonstrated
    in the design of nanophotonic fiber couplers,
    which couple light from an optical fiber at normal incidence
    into an in-plane waveguide\cite{baets}.

The structure of the optical fibers used was a 2 micron diameter core with 
    refractive index $n_\text{core} = 1.6$,
    surrounded by a cladding with refractive index $n_\text{cladding}=1.5$.
The reduced size of the fiber core was employed in order
    to keep the device footprint small.
Additionally, the fiber coupler devices were only etched to half the membrane depth,
    in order to increase the asymmetry in the device structure.

\subsubsection{Compact fiber coupler}
We first present the design of a compact fiber coupler.
Such a device is said to be compact in that the functions
    of coupling into the plane, and focusing into a narrow waveguide
    are overlapped in the same device footprint.

Although the performance specification (\fig{3Db_fib_te})
    desired a coupling efficiency above 90\%,
    only 51.5\% efficiency was achieved (\fig{fib_te});
    however, this likely remains the highest efficiency demonstrated
    in a compact fiber coupler.

\myfig{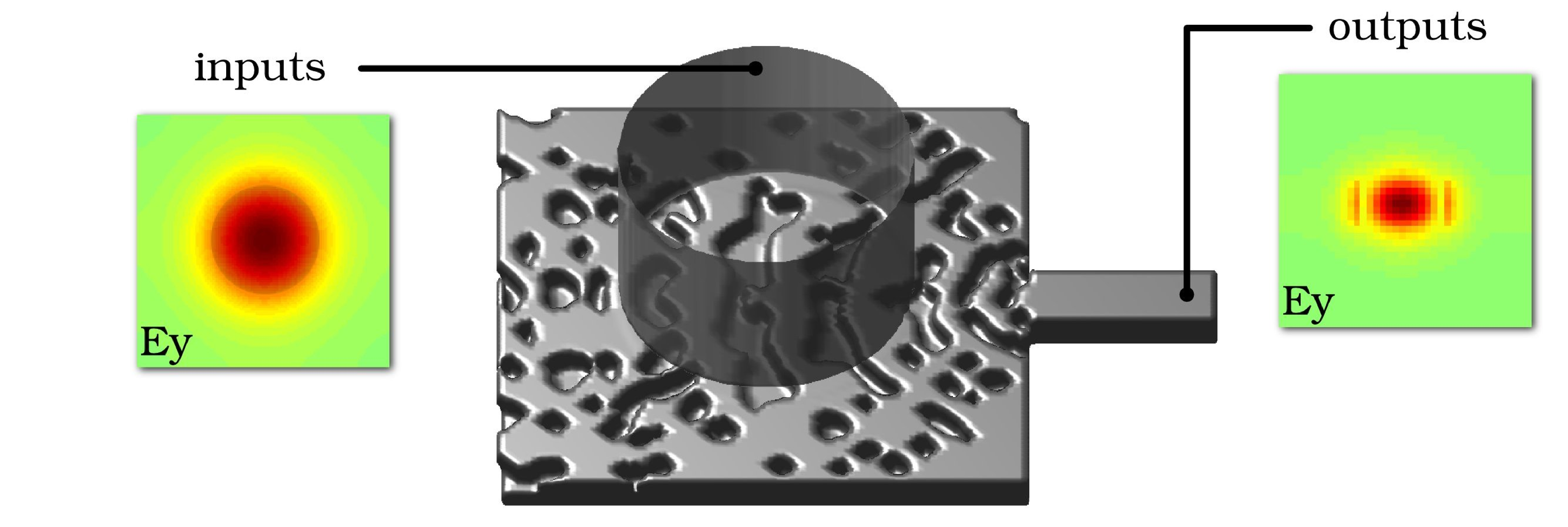}
    {Compact fiber coupler performance specification.
    Input mode consists of the $E_y$-polarized fundamental fiber mode.
    Output mode is the fundamental TE-polarized mode of the in-plane waveguide.
    Output power into the desired output arm is specified to be greater than 90\%.}
\myfig{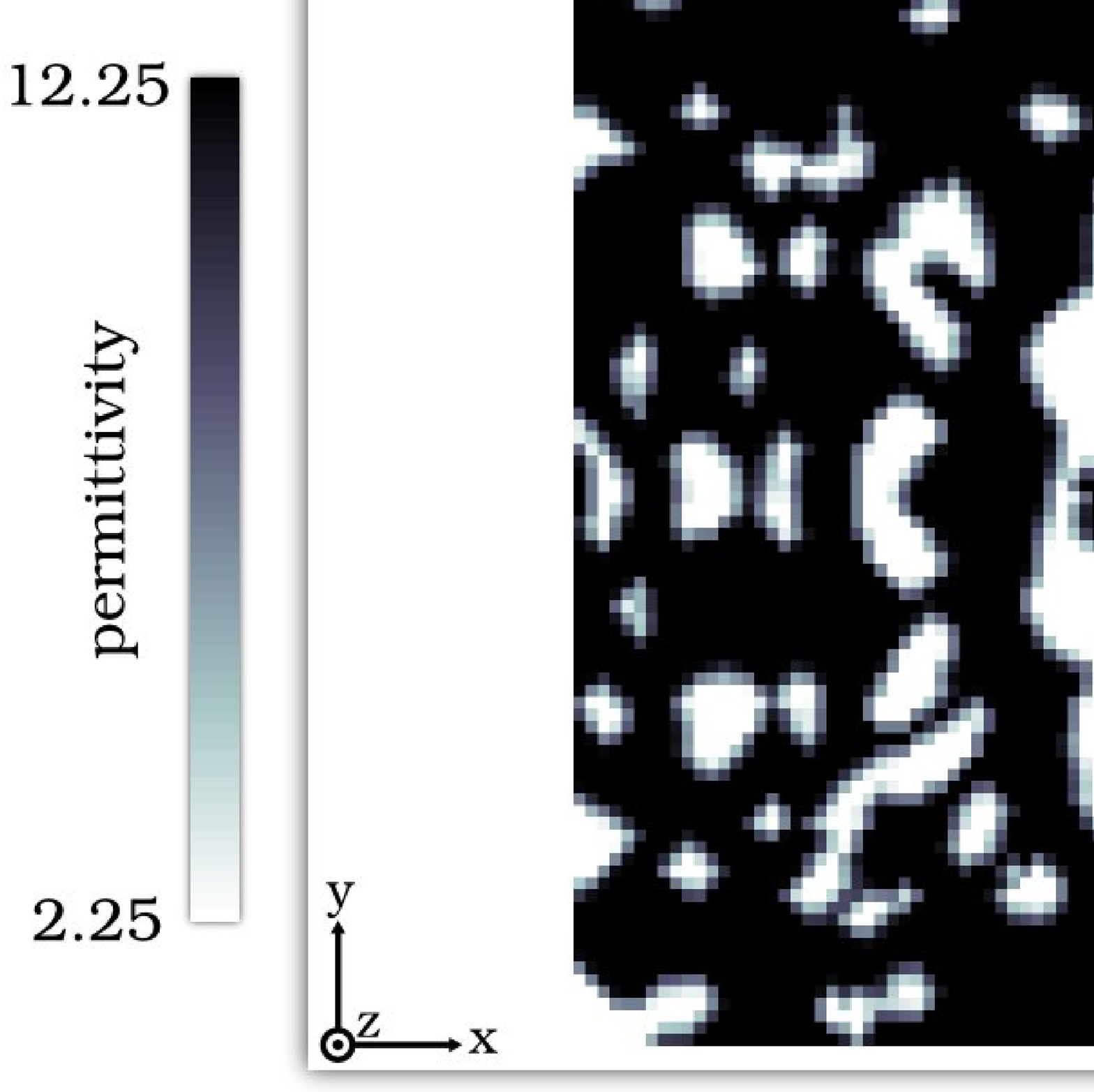}
    {Compact fiber coupler final result.
    The conversion efficiency into the in-plane waveguide mode is 51.5\%.}

\subsubsection{Mode-splitting fiber coupler}
We now continue to show how different functionalities
    can be incorporated into a single device,
    by virtue of our design-by-specification scheme.

Here, we show how the functionality of a fiber coupler
    can be combined with that of a spatial mode splitter.
Specifically, the performance specification (\fig{3Db_fib_pol})
    determines that different fiber spatial modes
    by split into different in-plane nanophotonic waveguides.

The final result (\fig{fib_pol}) has lower efficiencies;
    however, the result is still useful 
    in that no device with such a functionality has previously been demonstrated.

\myfig{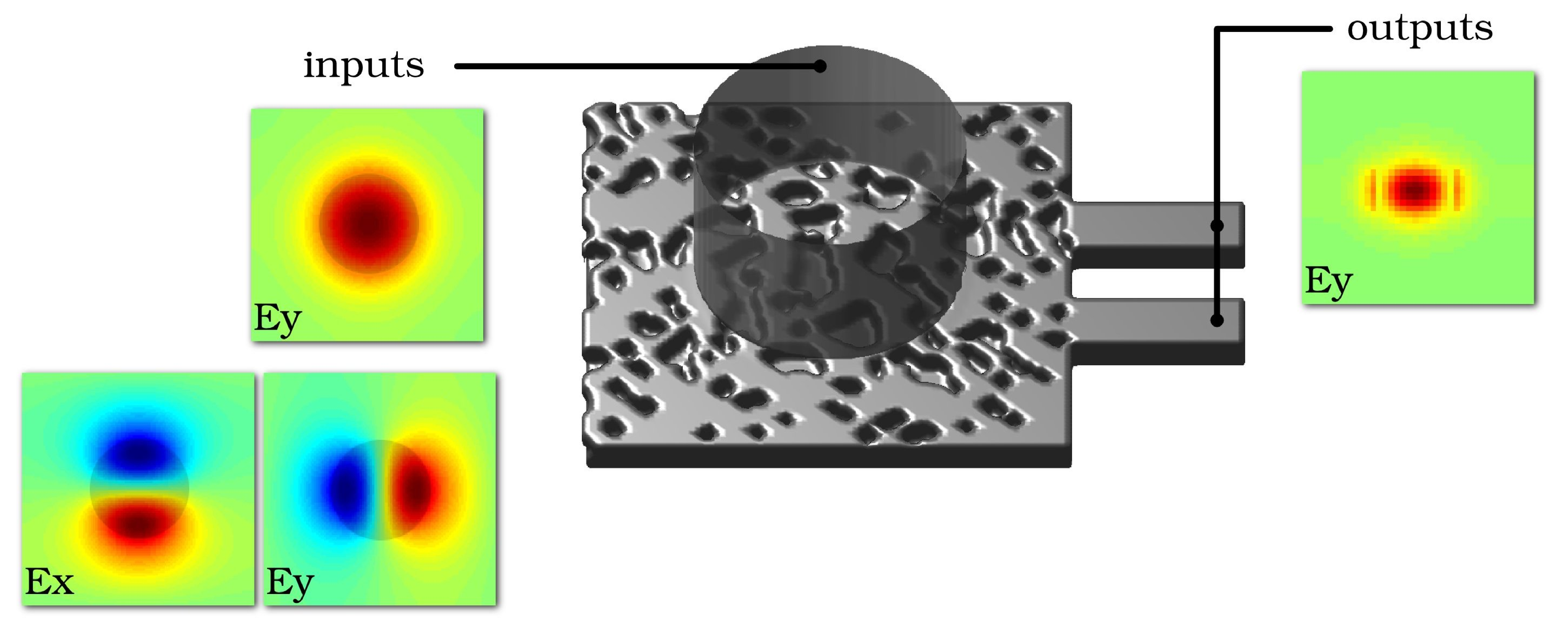}
    {Mode-splitting fiber coupler performance specification.
    Input mode consists of the fundamental fiber mode or 
        the third-order, circularly polarized fiber mode.
    Output mode is the fundamental TE-polarized mode of either in-plane waveguide.
    The device is designed to couple the fundamental fiber mode into
        the upper output arm, 
        while the third-order fiber mode is coupled into the lower output arm.
    Output power into the desired output arm is specified to be greater than 90\%.}
\myfig{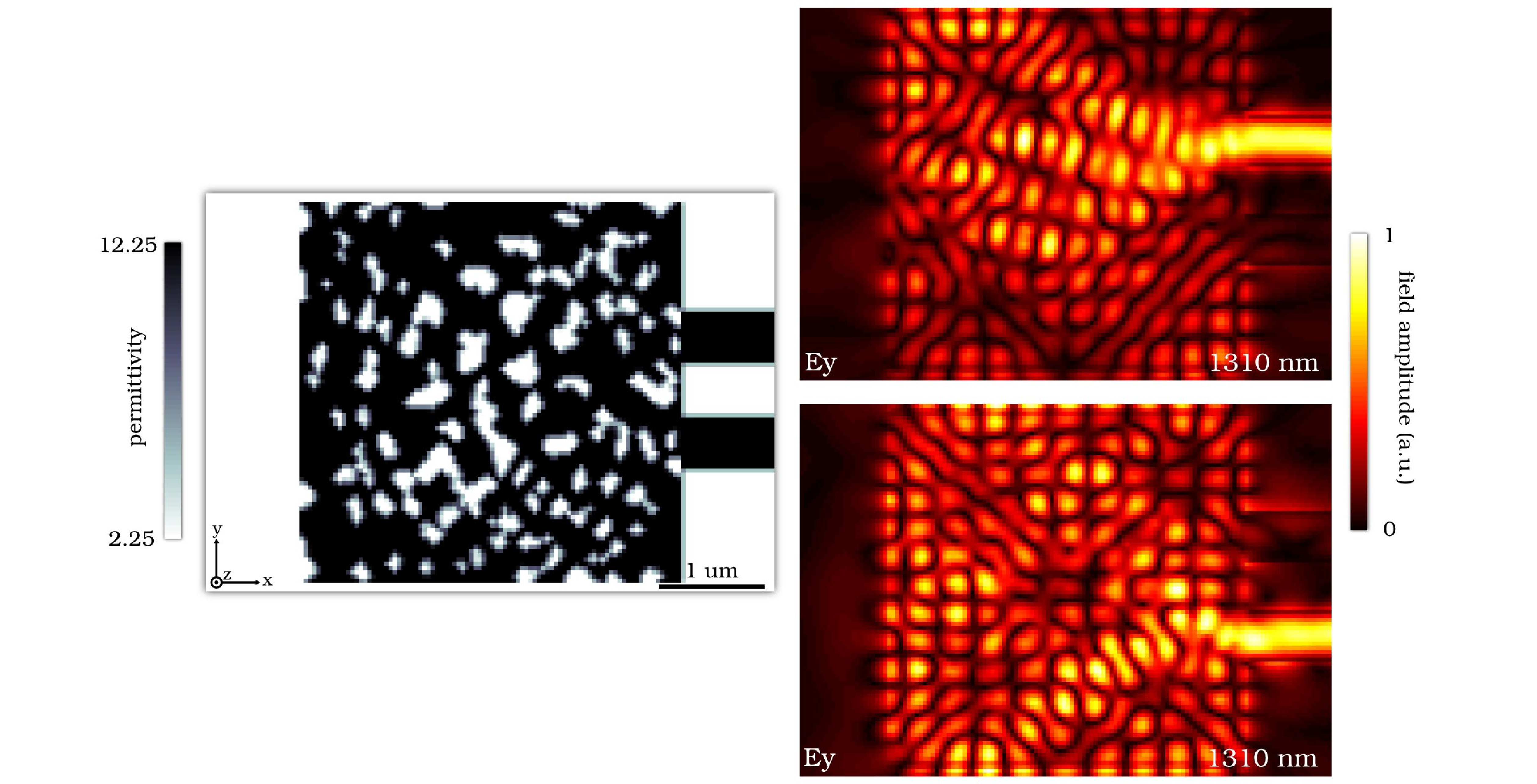}
    {Mode-splitting fiber coupler final result.
    The conversion efficiency for the fundamental fiber mode input is 32.6\% 
        (top plot).
    The conversion efficiency for the third-order fiber mode input is 22.7\% 
        (bottom plot).}

\subsubsection{Wavelength-splitting fiber coupler}
Another example of a functionality-combining device is the wavelength-splitting
    fiber coupler.
Here, fiber modes of different wavelengths are coupled in-plane
    and then split into different nanophotonic waveguides (\fig{3Db_fib_wdm}).
Once again, efficiencies are low (\fig{fib_wdm}),
    but no such device has previously been demonstrated.

\myfig{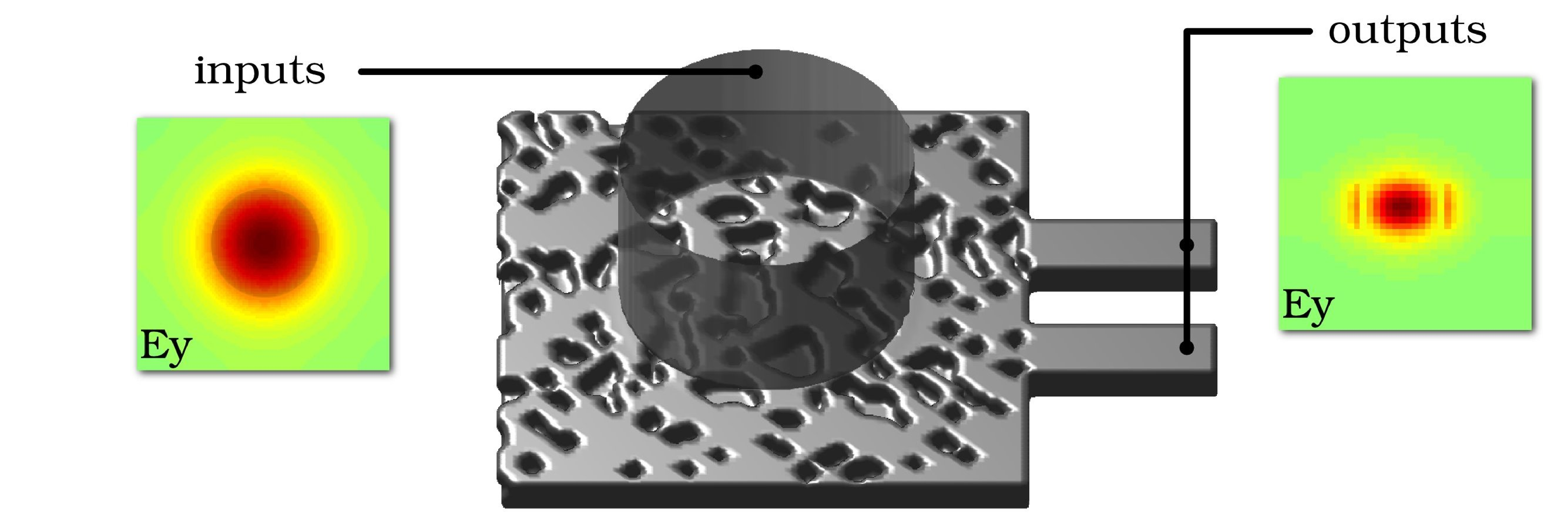}
    {Wavelength-splitting fiber coupler performance specification.
    Input mode consists of the $E_y$-polarized fundamental fiber mode at either
        the 1310 nm or 1550 nm wavelengths.
    Output mode is the fundamental TE-polarized mode of either in-plane waveguide.
    The structure is designed to guide light at the 1550 nm wavelength
        into the upper output arm,
        while 1310 nm light is guided into the lower output arm.
    Output power into the desired output arm is specified to be greater than 90\%.}
\myfig{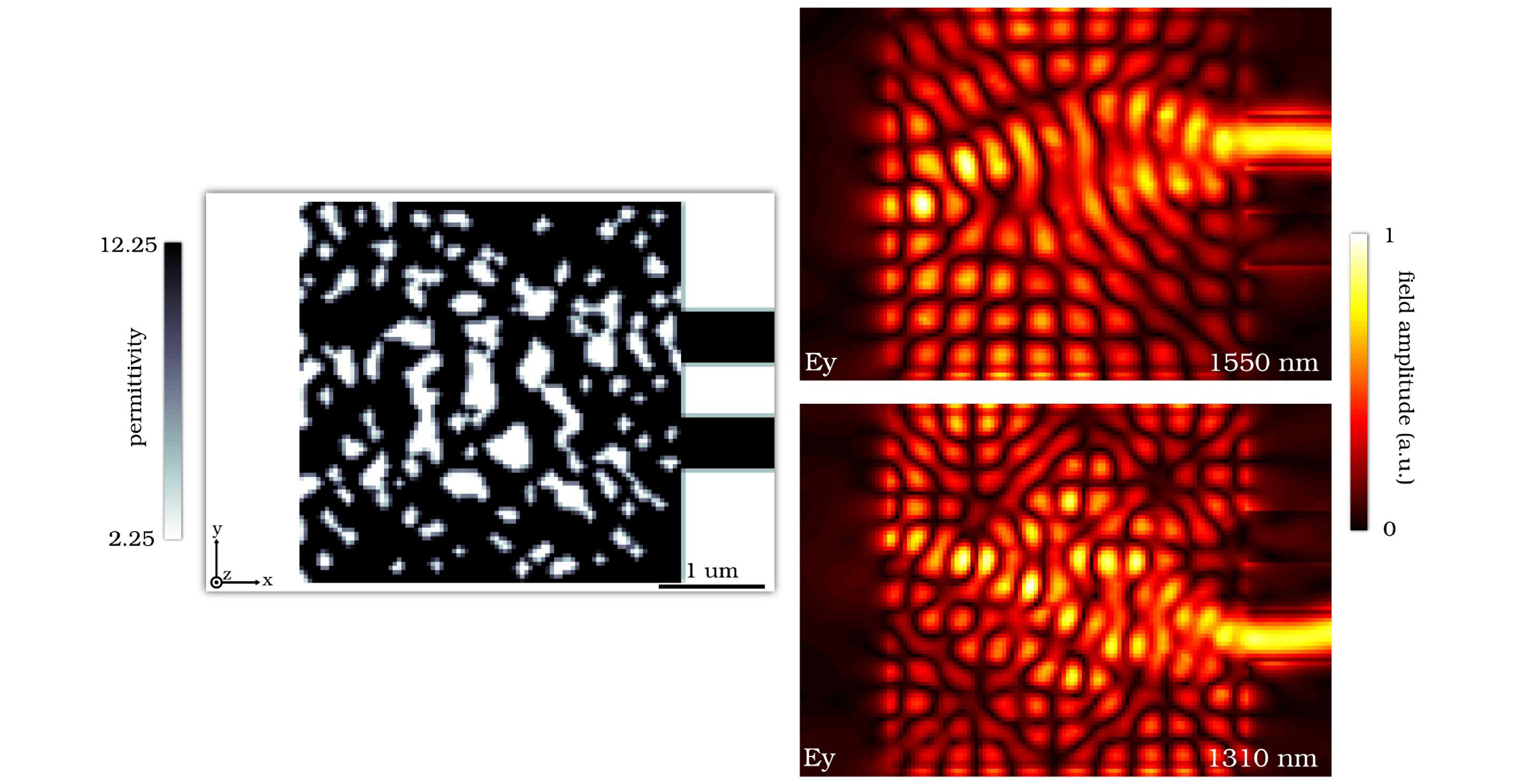}
    {Wavelength-splitting fiber coupler final result.
    The conversion efficiency into the in-plane waveguide mode at 1550 nm is 31.6\%.
    The conversion efficiency into the in-plane waveguide mode at 1310 nm is 28.6\%.}

\subsection{Broadband wavelength splitter}
We continue to investigate the capabilites of our method
    by attempting the design of a broadband wavelength splitter.

First, we revisit our wavelength splitter result (\fig{spl_wdm})
    and perform a broadband analysis, 
    the results of which are shown in \fig{analysis_spl_only}.
This analysis reveals that device performance quickly drops off
    as one moves away from the target wavelengths.
\myfig{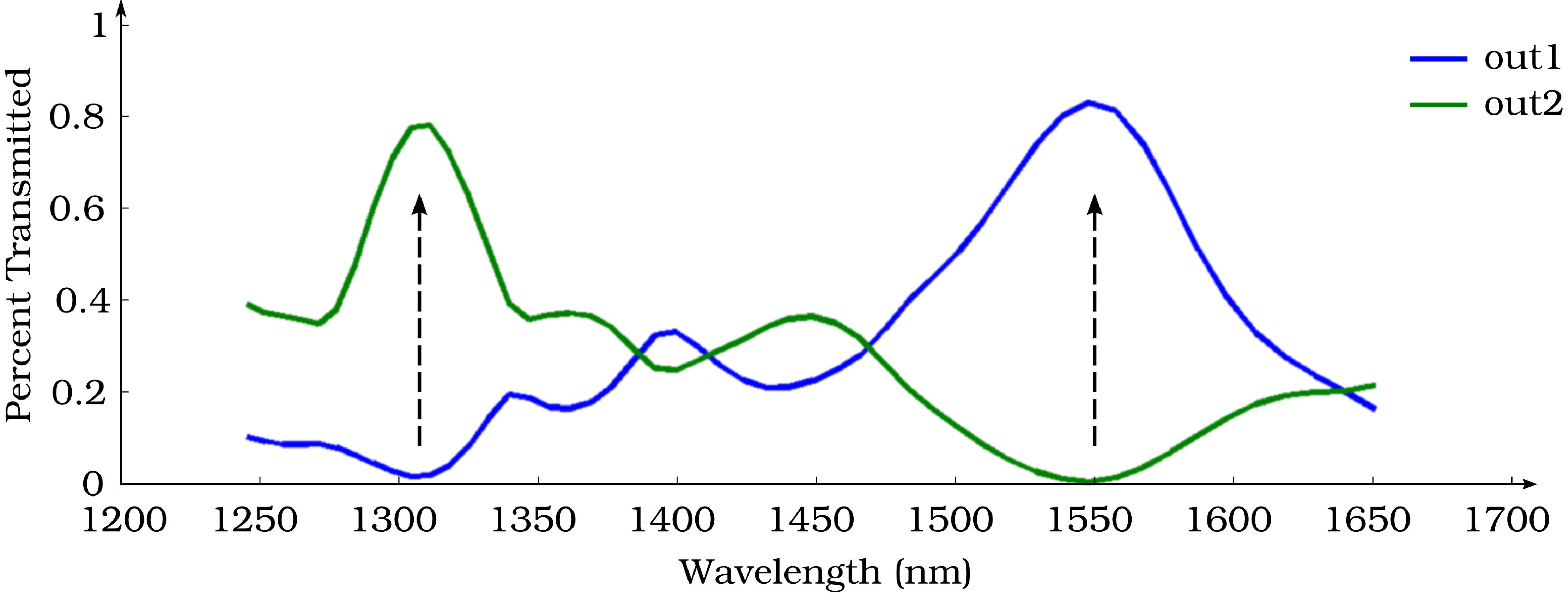}
    {Broadband analysis of previously design wavelength splitter (\fig{spl_wdm}).
    Although high-efficiency operation is achieved, 
    the performance quickly drops off away from the target wavelengths 
    (denoted by arrows).}

In order to design a broadband wavelength splitter,
    we modify our performance specification to include multiple target 
    wavelengths (with identical desired performance) around the 
    original target wavelengths, as seen in \fig{analysis_top}
    which reveals that broadband operation has been achieved.
The final result for the broadband wavelength splitter is shown in \fig{top_wdm}.
\myfig{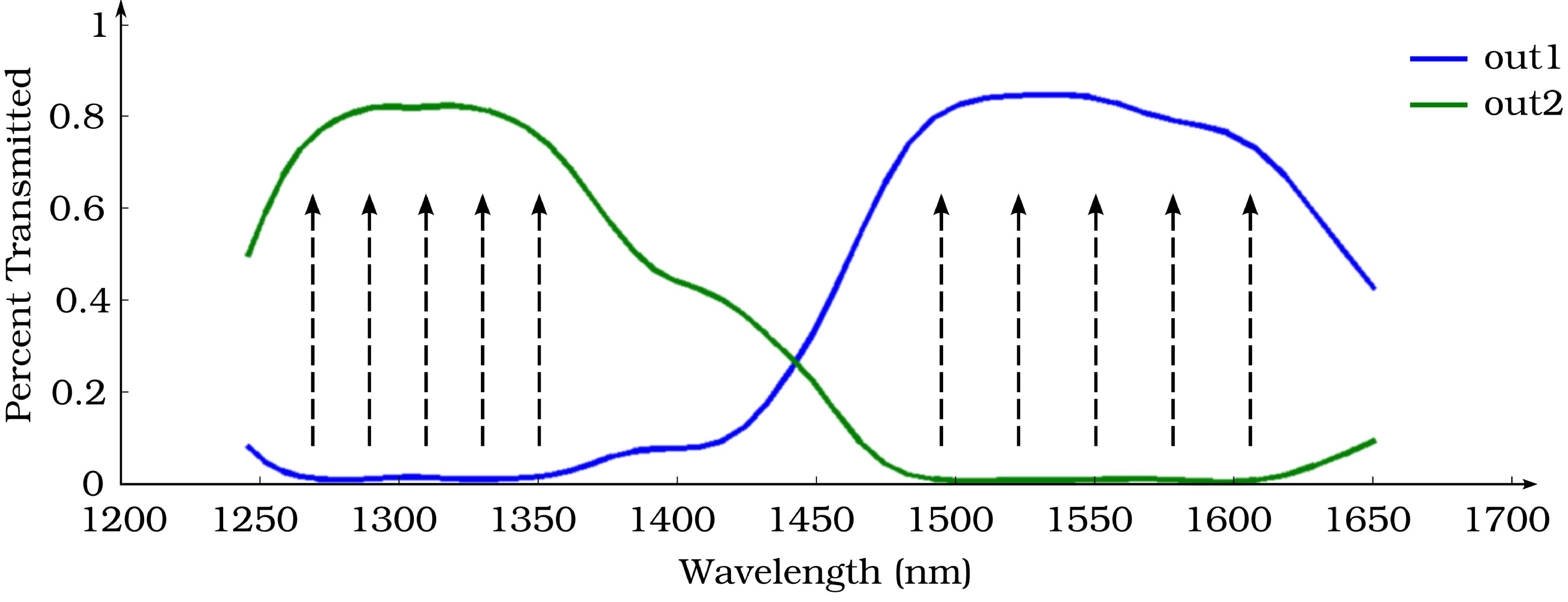}
    {Broadband analysis of broadband wavelength splitter 
        (final design shown in \fig{top_wdm}.
    The addition of multiple target wavelengths (vertical arrows)
    allows for 
    high-efficiency operation is achieved across a wide bandwidth.}
\myfig{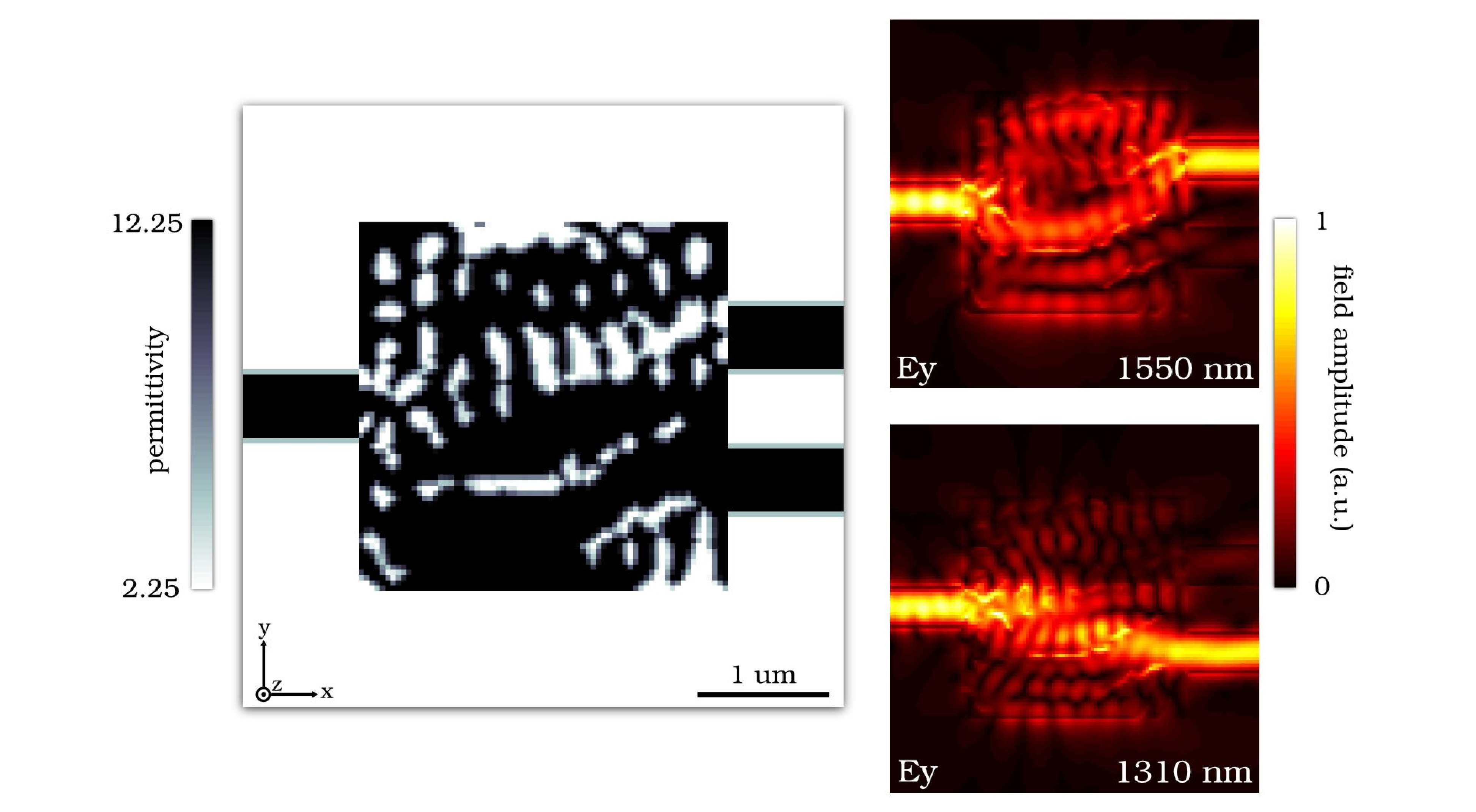}
    {Broadband wavelength splitter final result.
    The efficiencies at the central target wavelengths of 1550 nm and 1310 nm
    exceed those of its narrowband counterpart (\fig{spl_wdm}).}

\subsubsection{Temperature-robustness of broadband wavelength splitter}
We can now perform a temperature analysis of our broadband wavelength splitter,
    using $\Delta n_\text{Si} / \Delta T = 1.85 \cdot 10^{-4} K^{-1}$ 
    and $\Delta n_{\text{SiO}_2} / \Delta T = 0$ 
    (no refractive index shift for silica).
This analysis, shown in \fig{analysis_temp_shift2},
    reveals that stable operating points exist
    over a temperature range of nearly 1000 K.
Such a result is telling in that it demonstrates
    that on-chip optical devices can be designed to be \emph{passively} stable
    to temperature shifts which would typically be present in CPUs,
    since these are much less than 1000 K.
\myfig{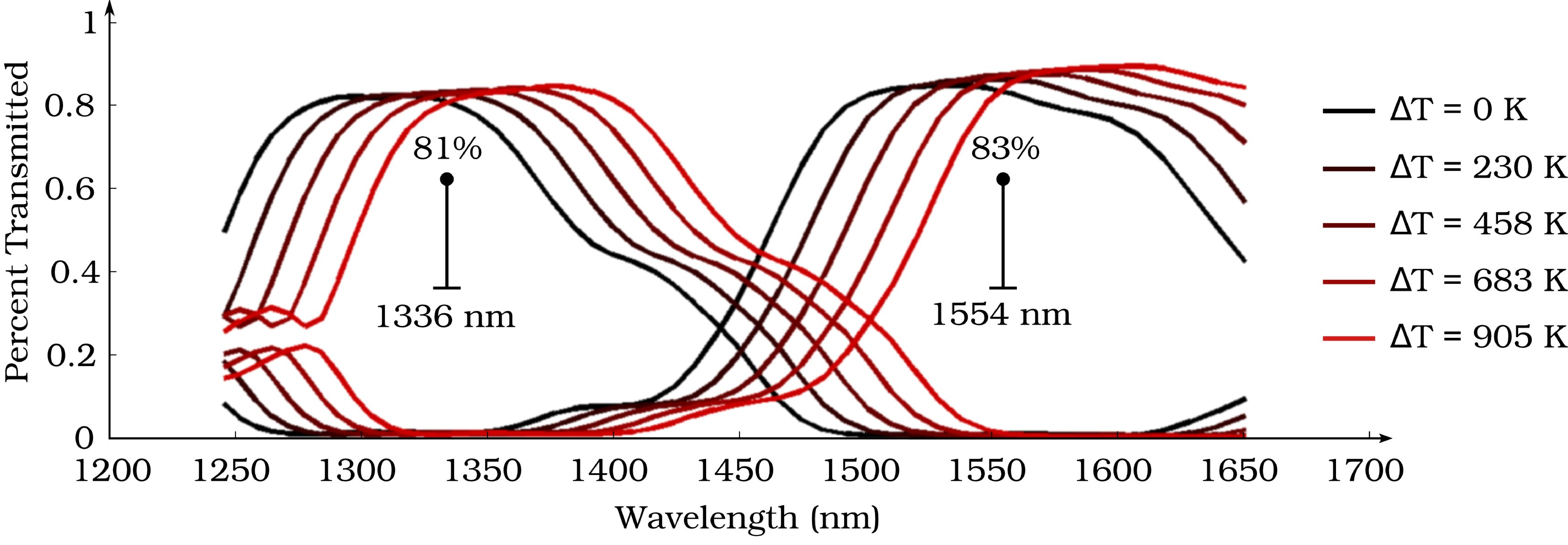}
    {Temperature analysis of the broadband wavelength splitter.
    Stable operating points (defined as efficiency $\ge$ 80\%)
    exist over a temperature shift of 905K.}

\subsubsection{Fabrication-robustness of broadband wavelength splitter}
A analysis with regard to fabrication-error
    was also performed on the broadband wavelength splitter.
The specific fabrication error was a general over- or under-etch
    of the device (input and output waveguides unaffected).
\Fig{analysis_fab_shift2} reveals that up to 8 nm of over- or under-etching
    can be sustained before performance falls below 70\%, at the central
    operating wavelengths.
The structural variations at 8 nm of etch error are shown in \fig{top_wdm_overunder}.
    
\myfig{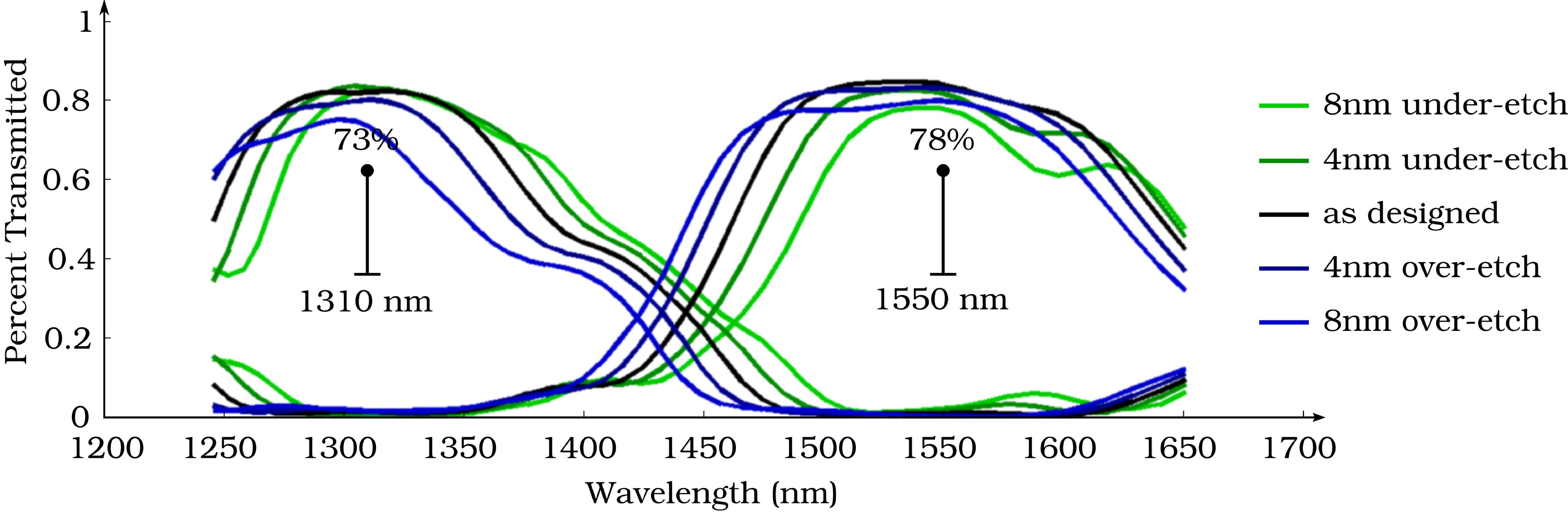}
    {Analysis of fabrication-error on the performance of the broadband 
    wavelength splitter.
    Original central wavelengths are shown to hold greater than 70\% efficiency,
    in spite of up to 8 nm of over- or under-etch error.}
\myfig{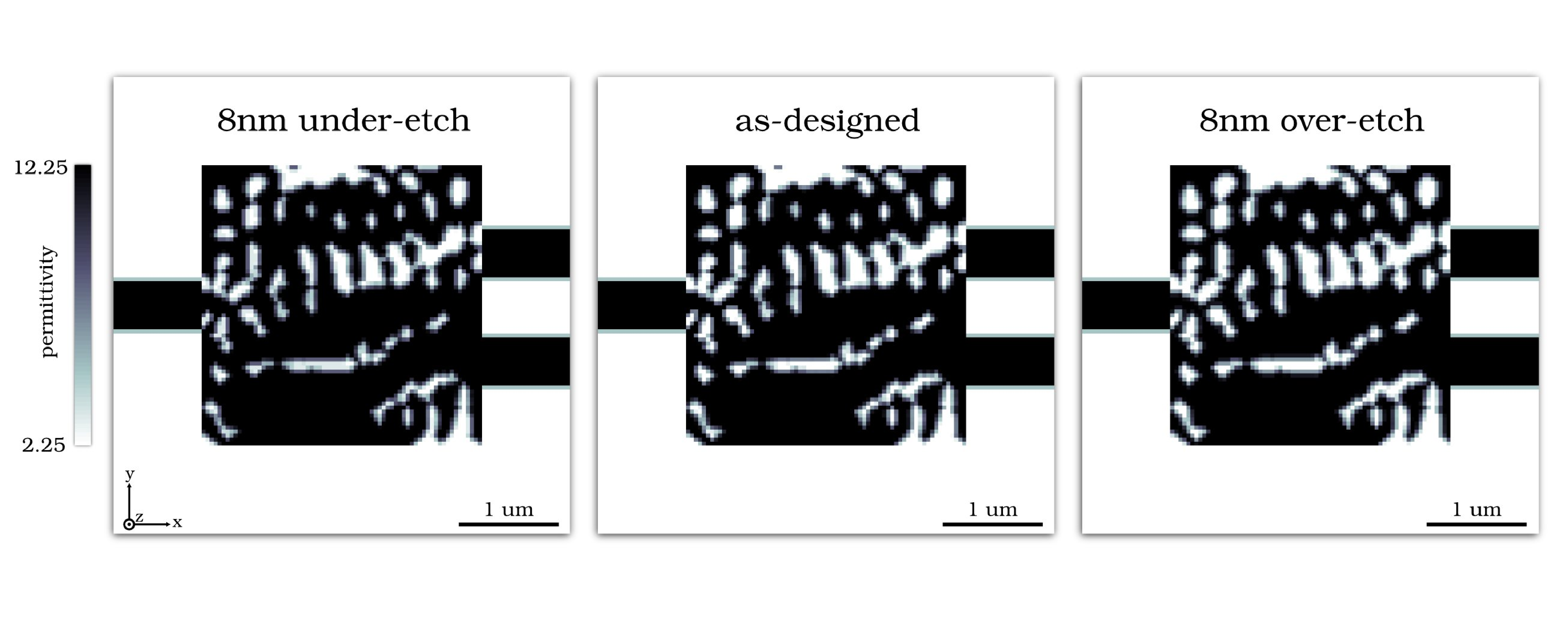}
    {Comparison of under-etched, as-designed, and over-etched structures.
    Differences are subtle since the pixel size is 40 nm and the
    fabrication error is 8 nm.}

This result is significant in that it demonstrates
    that the design of broadband devices
    seems to be a valid heuristic in the search for 
    devices which are tolerant to temperature shifts and fabrication error.
Note, however, that our method, as formulated, is also able to
    deal with temperature and fabrication shifts explicitly as well,
    although such results are not demonstrated here.

\section{Conclusion}
We have developed and implemented a method to design linear nanophotonic structures
    which are fully three-dimensional and multi-modal,
    have very compact footprints,
    exhibit high efficiency, and
    are manufacturable.
We demonstrate this capability by designing various nanophotonic mode converters,
    splitters, hubs, and fiber couplers.
Critically, many, if not all, of these devices have never been demonstrated before
    and cannot be designed by hand.
In contrast, our method allows user to easily design such devices
    by virtue of our design-by-specification scheme.

In addition, we demonstrate the design of a broadband device 
    which is strongly robust to wavelength and temperature shift,
    as well as fabrication error.
We show that such a device has stable operating wavelengths 
    over temperature shifts as large as 905 K,
    or over-/under-etching error of up to 8 nm.
We suggest, based on this design, that wavelength tolerance
    may be a good heuristic to the design of temperature and fabrication-error
    tolerant nanophotonic devices. \\

This work has been supported by the 
    AFOSR MURI for Complex and Robust On-chip Nanophotonics 
    (Dr. Gernot Pomrenke), grant number FA9550-09-1-0704.
\end{document}